\documentclass[aps,prl,superscriptaddress,showpacs,twocolumn,nofootinbib]{revtex4-1}
\usepackage[pdftex]{graphicx}
\usepackage{amsmath}
\usepackage{xspace}
\usepackage{color}
\usepackage[colorlinks=true,linkcolor=blue,urlcolor=blue,citecolor=blue]{hyperref}
\usepackage[normalem]{ulem}
\usepackage{hyperref}

\usepackage{mathrsfs}

\usepackage{color}
\definecolor{BV}{rgb}{0.1,0.,0.6}
\definecolor{R}{rgb}{0.9,0,0}
\definecolor{G}{rgb}{0.2,0.8,0.2}
\usepackage{comment}

\newcommand{\be}{\begin{equation}}
\newcommand{\ee}{\end{equation}}

\begin{document}

	\title{Nonequilibrium phase transition \\ in transport through a driven quantum point contact}

	\author{Oleksandr Gamayun}
	\email{o.gamayun@uva.nl}

	\affiliation{%
		Institute of Physics and Delta Institute for Theoretical Physics, University of Amsterdam
		\\
		Postbus 94485, 1090 GL Amsterdam, The Netherlands
	}
	\affiliation{Bogolyubov Institute for Theoretical Physics \\ 14-b Metrolohichna str. Kyiv, 03143, Ukraine}
	\author{Artur Slobodeniuk}
	\affiliation{
		Department of Condensed Matter Physics, Faculty of Mathematics and Physics, Charles University,
		\\ Ke Karlovu 5, Praha 2 CZ-121 16, Czech Republic
	}%

	\author{Jean-S{\'e}bastien Caux}%

	\affiliation{%
		Institute of Physics and Delta Institute for Theoretical Physics, University of Amsterdam
		\\
		Postbus 94485, 1090 GL Amsterdam, The Netherlands
	}%

	\author{Oleg Lychkovskiy}
	\affiliation{%
		Skolkovo Institute of Science and Technology\\
		Bolshoy Boulevard 30, bld. 1, Moscow 121205, Russia
	}%
	\affiliation{Laboratory for the Physics of Complex Quantum Systems, Moscow Institute of Physics and Technology, Institutsky per. 9, Dolgoprudny, Moscow  region,  141700, Russia
	}
	\affiliation{Department of Mathematical Methods for Quantum Technologies,
		Steklov Mathematical Institute of Russian Academy of Sciences\\
		8 Gubkina St., Moscow 119991, Russia
	}

	\date{\today}

	%%%%%%%%%%%%%%%%%%%%%%%%%%%%%%%%%%%%%%%%%
	%%%%%%%%%%%%%%%%%%%%%%%%%%%%%%%%%%%%%%%%%
\begin{abstract}
We study transport of noninteracting fermions through a periodically driven quantum point contact (QPC) connecting two tight-binding chains. Initially each chain is prepared in its own equilibrium state, generally with a bias in chemical potentials and temperatures. We examine the heating rate (or, alternatively,  energy increase per cycle) in the nonequilibrium time-periodic steady state established after initial transient dynamics. We find that the heating rate vanishes identically when the driving frequency exceeds the bandwidth of the chain. We first establish this fact for a particular type of QPC where the heating rate can be calculated analytically. Then we verify numerically that this nonequilibrium phase transition is present for a generic QPC. Finally, we derive this effect perturbatively in leading order for cases when the QPC Hamiltonian can be considered as a small perturbation. Strikingly,  we discover that for certain QPCs the  current averaged over the driving cycle also vanishes above the critical frequency, despite a persistent bias. This shows that a driven QPC can act as a frequency-controlled quantum switch.
\end{abstract}

	\pacs{02.30.Ik,05.70.Ln,75.10.Jm}

	\maketitle

	%%%%%%%%%%%%%%%%%%%%%%%%%%%%%%%%%%%%%%%%%
	%%%%%%%%%%%%%%%%%%%%%%%%%%%%%%%%%%%%%%%%%

%\noindent{\it Introduction.}~
Controlling the state of the electron gas by means of external time-dependent potentials is one of the central challenges of condensed matter physics with immediate applications to micro- and nanoelectronics. Of particular interest is transport through quantum point contacts (QPC) with time-dependent parameters.
Many remarkable phenomena have been predicted and observed in driven QPCs,
ranging from quantum pumps \cite{PhysRevLett.67.1626,Nakajima2016topological,Lohse2015thouless} to noise-free excitation of particles from the Fermi Sea \cite{Levitov_1996,Ivanov_1997,Keeling_2006,Dubois_2013}.
On the practical side, the creation of new electronic devices suitable in particular for quantum computation remains an alluring prospect. For instance, time-dependent QPCs can be considered as a means to ``braid'' Majorana fermions in topological superconductors \cite{Nayak_2008,Beenakker_2019}.
Theoretical approaches to these problems include the adiabatic modification of the Landauer-B\"uttiker formalism for slow drives \cite{Moskalets_2002,PhysRevB.68.155304,PhysRevB.71.045326,Sherkunov_2010,Moskalets_2011},  Keldysh perturbation theory \cite{PhysRevLett.99.076601} and various approximation schemes based on Floquet theory and the theory of open quantum systems \cite{KOHLER_2005,PhysRevB.69.165306,PhysRevLett.116.026801}.

 % central theme of applied research in micro- and nano-electronics, as well as fundamental studies of quantum out-of-equilibrium physics Floquet physics and one of the main directions of research in nanoelectronics. 	In quantum transport experiments it is remarkably manifested in various physical phenomena ranging from quantum pumps to a noise-free excitations of the particles from the Fermi Sea \cite{Keeling_2006}. 	The practical interest lies in the posibility to realize the quantum computation, for instance, by 	"braiding" of the Majorana fermions in topological superconductors \cite{Nayak_2008,Beenakker_2019}.

\begin{figure}[t] %  figure placement: here, top, bottom, or page
		\centering
\includegraphics[width=\linewidth]{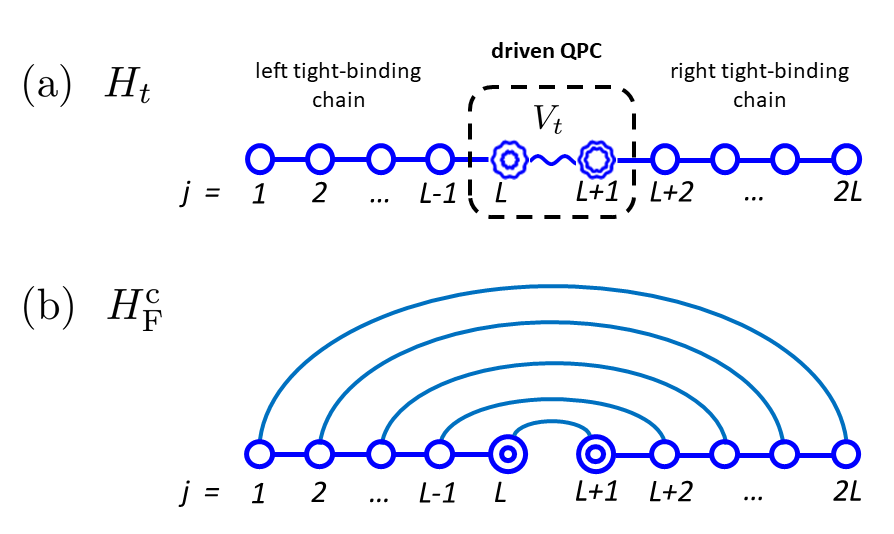}
        \caption{(a) Driven QPC connecting two tight-binding chains. On-site potentials indicated by double lines can be present on two QPC sites. Wiggly lines indicate time-dependence of the QPC. The whole system is described by the Hamiltonian~\eqref{Ht} (b) The Floquet Hamiltonian~\eqref{HFl} of the system with the conformal QPC~\eqref{conformal QPC}. }
		\label{Fig:Hconf}
\end{figure}

Here we revisit transport through a periodically driven QPC in a simple setting of noninteracting fermions. Namely, we consider a closed quantum system consisting of two one-dimensional tight-binding chains connected by a QPC. The latter is described by a periodic time-dependent potential $V_t$ with a period $\tau$. We assume that it acts nontrivially only on adjacent edge sites of the two chains, see Fig. \ref{Fig:Hconf} (a).

We assume that initially each chain is in its own equilibrium, possibly with different particle densities and temperatures.
%For presentational purposes, we mainly focus on the case when the left chain is initially filled with fermions at equilibrium with respect to the Hamiltonian $H_L$, while the right chain is empty.\footnote{All our results are also applicable to more general initial states with both chains filled with fermions, see below.
%% in equilibrium states filled by fermions in equilibrium , which is a direct product of two equilibrium  with both chains filled with fermions in equilibrium with respect  straightforwardly generalized to more general initial conditions, as  below and in  the Supplement \cite{supplementary}.
%}
One generally expects that in such setting a nonequilibrium time-periodic steady state will be established in the vicinity of the QPC after initial transient dynamics.
% similarly to what happens for a static QPC \cite{Gamayun_2020,Ljubotina_2019,Doyon_2019}.
%\footnote{It should be emphasized that we always think of system sizes large enough to avoid finite-size effects (such as reflections from the distant edges of the chains) on this steady state.}
We focus on two quantities characterizing this steady regime: the heating rate $\overline{\mathscr{W}}$ and current  through the QPC,  $\overline{\mathcal{J}}$, both averaged over the driving period $\tau$. We consider system sizes large enough to avoid any finite-size distortions.
%The energy the energy increase per cycle, $\Delta \mathcal{E}$, and the number of fermions (charge) $\overline{\mathcal{J}}$  transferred per driving cycle through the QPC from left to right.
%Note that $\tau \overline{\mathscr{W}} $ is the increase of the total energy per driving cycle, while  $\tau \overline{\mathcal{J}}$ is the number of fermions transferred per cycle through the QPC from left to right.
The first main result of the present manuscript is that  the  heating rate $\overline{\mathscr{W}}$ experiences a {\it nonequilibrium phase transition}\footnote{Nonequilibrium phase transition
refers to a singular behavior of observables in the nonequilibrium
steady state as a function of control parameters \cite{marro2005nonequilibrium,tauber2014critical,prosen2011nonequilibrium}. Specifically, in our
case the observables are the heating rate and the current, and the
control parameter is the driving frequency.}
for an arbitrary QPC, vanishing identically when the frequency  of the drive, $\omega=2\pi/\tau$, exceeds a critical value equal to the single-particle bandwidth of the chain. An analogous effect, but for a  \textit{global} driving, was found in a spin system \cite{dalessio2013many-body} and in a system of coupled Kapitza pendulums  \cite{citro2015dynamical}, where it was interpreted as an energy localization transition. The second main result is that for some  $V_t$  the current $\overline{\mathcal{J}}$ also vanishes above the critical frequency, despite a finite difference in particle densities and temperatures between the chains. Given that at almost any moment of time there is a nonzero tunneling matrix element connecting two chains,  this latter finding seems particularly counterintuitive.
%This effect is similar to the dynamical localization in real space \cite{dunlap1986dynamic}  and the coherent destruction on tunneling \cite{grossmann1991coherent,reyes2017transport,agarwala2017effects}.
We discuss similarities and differences between our results and relevant prior work \cite{dalessio2013many-body,citro2015dynamical,dunlap1986dynamic,grossmann1991coherent,reyes2017transport,agarwala2017effects,russomanno2012periodic} in the end of the manuscript.

We substantiate our claims in three ways. First, we consider a particular  QPC -- a {\it conformal} QPC -- which allows for a completely analytical treatment. We explicitly construct the Floquet Hamiltonian and calculate   $\overline{\mathscr{W}}$ and   $\overline{\mathcal{J}}$. It turns out that in this case  $\overline{\mathscr{W}}$ vanishes above the critical frequency while $\overline{\mathcal{J}}$ remains finite.

We subsequently numerically examine various QPCs. It is observed that $\overline{\mathscr{W}}$ generically experiences a phase transition, while $\overline{\mathcal{J}}$ -- only does so for certain QPCs.

Finally, for a small $V_t$ we calculate  $\overline{\mathscr{W}}$ and  $\overline{\mathcal{J}}$ in leading order of perturbation theory, where we confirm the universal nature of the phase transition of $\overline{\mathscr{W}}$ and elucidate one of the conditions for the phase transition of  $\overline{\mathcal{J}}$.

%Our findings are related to prior work on the quantum Floquet dynamics, most importantly, on dynamical localization in real \cite{dunlap1986dynamic} and energy \cite{dalessio2013many-body,citro2015dynamical} spaces and on coherent destruction of tunneling \cite{grossmann1991coherent,reyes2017transport,agarwala2017effects}. We discuss similarities and differences between our results and relevant prior work in the end of the manuscript.

\medskip
\noindent{\it General setup.}~ The total Hamiltonian of the system is
\be\label{Ht}
H_t=H_L+H_R+V_t,
\ee
where $H_L$  and $H_R$ describe two tight-binding chains that are disconnected in the absence of the QPC,
%\begin{equation}
%H_L+H_R = - \frac{1}{2}\sum_{j=1}^{L-1} c_j^\dagger c_{j+1} -\frac{1}{2}\sum_{j=L+1}^{2L-1}  c_j^\dagger c_{j+1} + {\rm h.c. },
%\end{equation}
\begin{align}
H_{L}&=-\frac12 \sum_{j=1}^{L-1} (c_j^\dagger c_{j+1}+ c_{j+1}^\dagger c_j),\nonumber\\
H_{R}&=-\frac12 \sum_{j=L+1}^{2L-1} (c_j^\dagger c_{j+1}+ c_{j+1}^\dagger c_j),
\end{align}
where $c^\dagger_j$ ($c_j$) are creation (annihilation) fermionic operators. The single-particle spectrum of each chain is given by $E_p=-\cos p$, where $p\in[0,\pi]$ is the quantized quasimomentum, and the single-particle energy bandwidth  is equal to $2$.
%The bandwidth of each chain is equal to 2.
%$\mathscr{E}(k)=-\cos k$,

%$$\mathcal{J}\qquad\mathscr{J}\mathcal{W}\qquad\mathscr{W}$$

The QPC  is described by
% describes the driven QPC. It acts nontrivially only on two adjacent edge sites of two chains, see Fig. \ref{Fig:Hconf} (a), and is given by
\begin{equation}\label{VTT}
V_t = - \frac{1}{2} \left(c^\dagger_L\,c_{L+1}^\dagger\right) \left(\begin{array}{cc}
U_L & J_t\\
J_t & U_R
\end{array}\right)\left(
\begin{array}{c}
c_L \\ c_{L+1}
\end{array}
\right)
\end{equation}
%\be
%V_t=-\frac12 \left( J_t\, c_{L+1}^\dagger c_L+ J_t^*\,c_L^\dagger c_{L+1} + U^L_t\, c_L^\dagger c_L+U^R_t\,c_{L+1}^\dagger c_{L+1}  \right).
%\ee
Here  $J_t$, $U^{L,R}_t$  are real periodic functions of time  with a period $\tau$.
Physically, $J_t$ corresponds to the tunneling amplitude between the chains while $U^L_t$ and $   U^R_t$ are local on-site potentials (up to the the prefactor $-(1/2)$).
 %We assume that initially the chains are disconnected, i.e. $J_0=0$.
The whole system is illustrated in Fig.\ref{Fig:Hconf}(a). Initially each chain is separately prepared in its own equilibrium. This way, the initial state is characterised by the Fermi-Dirac occupation probabilities $\rho_L(E)$ and $\rho_R(E)$ of single-particle levels of the left and right chain, respectively.   %In particular, if at $t=0$ the left chain is in the ground state with $N_f$ fermions and the right chain is empty, then  $\rho_R(E)=0$ and $\rho_L(E)=\theta(-\cos k_F-E)$, where  $\theta(\dots)$ is  the step function and $k_F=\pi N_f/L$.

\medskip
\noindent{\it Conformal QPC.}~ We address analytically a driven conformal QPC defined by
\be\label{conformal QPC}
J_t=\sin \omega t,\quad U^L_t=-U^R_t=  \cos \omega t.
%V_t=V_t^{\rm c}=-\frac12 \sin \omega t \, (c_1^\dagger b_1+ b_1^\dagger c_1)-\frac12 \cos \omega t \, (b_1^\dagger b_1-c_1^\dagger c_1).
\ee
We refer to the Hamiltonian \eqref{Ht} with such parameters as $H_t^{\rm c}$.
A time-independent analog of this Hamiltonian  was  introduced in Ref.~\cite{Eisler_2012}. The transmission coefficient in  Ref.~\cite{Eisler_2012} is constant for all energies of the incoming particles (in  contrast to scattering on a generic defect), which resembles  the properties of the $S$-matrix  obtained by gluing together two conformal field theories \cite{Bachas_2002,Quella_2007}.

\begin{figure}[t] %  figure placement: here, top, bottom, or page
	\centering
	\includegraphics[width=\linewidth]{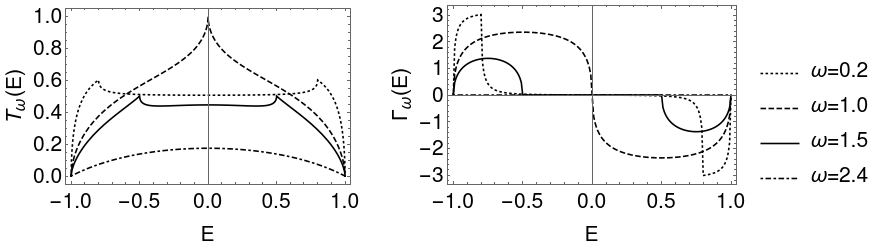}
	\caption{The transmission coefficient (left panel) and the heating function (right panel) for the conformal QPC for different driving frequencies.}
	\label{Fig:se}
\end{figure}

The major insight enabling a fully analytical treatment of the conformal QPC is that $H_t^{\rm c}$ can be represented
as a time-dependent unitary transformation of $H_0^{\rm c}$,
%
%The simple explanation of this fact is that the spectrum of the Hamiltonian
%$H_{t_0}^{\rm c}$ is $t_0$ independent, which means that at each moment of time $H_t^{\rm c}$ can be represented as a time-dependent unitary transformation of $H_0^{\rm c}$. We use this insight
%to get fully analytic treatment of the conformal QPC. Namely, we present
%%
%\footnote{ The term {\it conformal} is due to the fact that for a fixed $t$ the transmission coefficient in such model is independent on the energy of the incoming particle, thus resembling the properties of the S-matrix obtained by gluing together two conformal field theories \cite{Bachas_2002,Quella_2007}.}
%Note that at $t=0$ and, more generally, at $t=n \tau/2 $ ($n$ integer) two chains are disconnected.
%The major insight enabling the fully analytical treatment of the conformal QPC is that  $H_t^{\rm c}$ can be represented as a time-dependent unitary transformation of $H_0^{\rm c}$:
\be\label{Hconf unitary tranf}
H_t^{\rm c}=e^{i  \omega  t \Sigma/2} H_0^{\rm c}\, e^{-i \omega t \Sigma/2},\quad \Sigma=i\sum_{j=1}^L ( c_j^\dagger c_{2L+1-j}- {\rm h.c.}).
\ee
As a consequence, the solution of the Schr\"odinger equation $i\partial_t \Psi_t =H_t^{\rm c} \Psi_t$ can be recast in the form
\be
\Psi_t=e^{i  \omega  t \Sigma/2} e^{-i \left(H_0^{\rm c} +\omega  \Sigma/2\right)t}\Psi_0.
\ee
At  stroboscopic times $t_n$ (which are integers of the period, $t_n=n \tau$)   the first exponent reads $e^{i  \pi n\Sigma}=e^{i  \pi n N},$ where $N$ is the particle number operator. Therefore the stroboscopic evolution  is governed by
$
\Psi_{t_n}= e^{-i H^{\rm c}_{\rm F} t_n}\Psi_0
$, where the Floquet Hamiltonian $H^{\rm c}_{\rm F}$ reads
\be\label{HFl}
H^{\rm c}_{\rm F}=H_0^{\rm c} +\frac{\omega}2  \Sigma -\frac{\omega}2  N.
\ee
The last term does not affect the dynamics of particle-number-conserving quantities and is dropped henceforth. This Floquet Hamiltonian is illustrated in Fig.\ref{Fig:Hconf}(b).  Note that the term proportional to $\Sigma$ introduces long-range hoppings similar to \cite{arze2018out}.
This explicit form of $H^{\rm c}_{\rm F}$ allows us to perform full analysis of the driven dynamics, which can now be reduced to an equivalent quench dynamics.
In Supplemental Material \cite{supplementary} we show how to find analytical expressions for the averaged heating rate $\overline{\mathscr{W}}\equiv \Delta {\mathcal E}/\tau$ and current  $\overline{\mathcal{J}}\equiv \Delta N_R/\tau$, where $\Delta {\mathcal E}$ and $\Delta N_R$ are the increase  per driving cycle of the total energy and   the number of fermions in the right chain, respectively.\footnote{We emphasize that $\overline{\mathscr{W}}$ and $\overline{\mathcal{J}}$ characterize the nonequilibrium time-periodic steady state established  in the thermodynamically large system after initial relaxation. To calculate these  quantities we first take the limit $L\rightarrow \infty$ and then $t\rightarrow\infty$.} The result reads
\begin{equation}\label{dE}
\overline{\mathscr{W}} = \int \frac{dE}{2\pi\tau}\left(\rho_L(E)+\rho_R(E)\right)\Gamma_\omega(E),
\end{equation}
\begin{equation}\label{dQ}
\overline{\mathcal{J}} =  \int \frac{dE}{2\pi} \left(\rho_L(E)-\rho_R(E)\right)T_\omega(E).
\end{equation}
The transmission coefficient $T_\omega(E)$ and the heating function $\Gamma_\omega(E)$ can be found explicitly in elementary functions, see (S38), (S40), and are plotted in Fig.~\ref{Fig:se}.
The full time dependence can also be restored \cite{supplementary}.
%\begin{widetext}
%		\begin{multline}
%		\Gamma(E) =2\pi\frac{\sqrt{1-E^2}}{\omega^2} {\rm Re} \left[(\sqrt{1-(E-\omega)^2}(E^2 - \omega E -1)- \sqrt{1- (E+\omega)^2}(E^2 + \omega E-1))\right]+ \\
%		+ \pi\frac{1-E^2}{\omega^2} \left(
%		((E+\omega)^2-1) \theta(1- \omega-E) -
%		((E-\omega)^2-1) \theta(E- \omega+1)
%		\right).
%		\end{multline}
%	\end{widetext}
%The explicit expression for the transmission coefficient  $T_\omega(E)$ as well as the derivation of eqs. \eqref{dE},\eqref{dQ} can be found in the Supplement \cite{supplementary}.
\begin{figure}[t] %  figure placement: here, top, bottom, or page
		\centering
		\includegraphics[width=\linewidth]{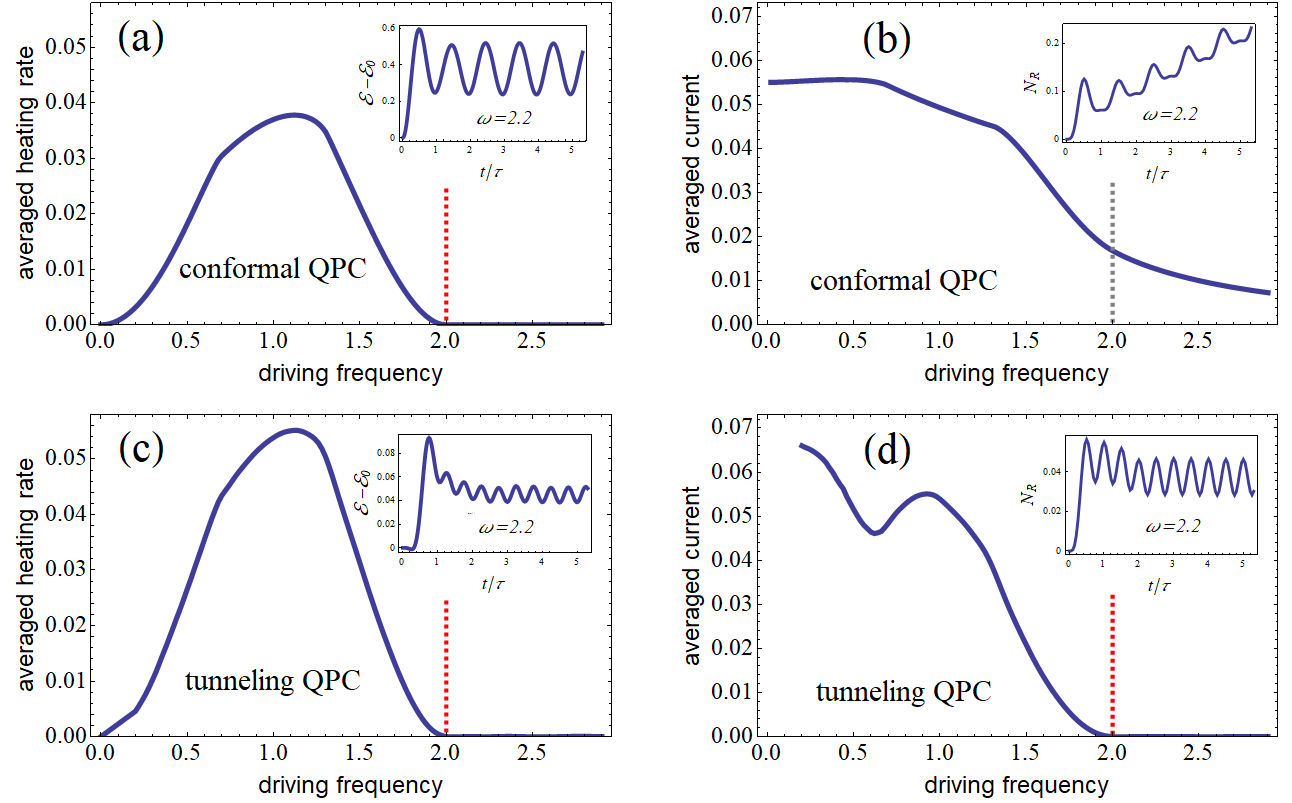}
		\caption{Average heating rate  $\overline{\mathscr{W}}$ (a,c) and average current through the QPC $\overline{\mathcal{J}}$ (b,d) {\it vs} driving frequency. Initially the left chain is filled by fermions in the ground state with particle density  $N/L=0.4$ fermions per site, while the right chain is empty. (a,b): Conformal QPC \eqref{conformal QPC}. The heating rate and the current are obtained analytically  from Eqs. \eqref{dE} and \eqref{dQ}, respectively. The heating rate vanishes for $\omega \geq 2$ while the current remains finite. (c,d): Tunneling  QPC \eqref{tunneling QPC}. Plots are obtained numerically for a system with $L=50$ sites in each chain and $N=20$ fermions. Both the heating rate and the current vanish for $\omega \geq 2$. Insets show the real-time dynamics  of  (a,c) the increase of the total energy, $\mathcal{E}-\mathcal{E}_0$,  and  (b,d) the particle number in the right chain, $N_R$.}
		\label{Fig:main}
\end{figure}
The non-smooth dependence of the transmission coefficient is associated with Floquet resonances (see below) and is present only in the thermodynamic limit.
The most remarkable feature, though, is that $\Gamma_\omega(E)$ turns to zero for $\omega\geq 2$, leading to
\begin{equation}\label{conf phase transition}
\overline{\mathscr{W}} =0 \quad {\rm for} \quad \omega \geq 2.
\end{equation}

We plot $\overline{\mathscr{W}}$ and $\overline{\mathcal{J}}$ as functions of $\omega$ in Fig. \ref{Fig:main} (a) and~(b), respectively. It can be seen that while $\overline{\mathscr{W}}$ experiences a phase transition at $\omega=2$ in accordance with eq. \eqref{conf phase transition}, this in not the case with $\overline{\mathcal{J}}$, meaning that some finite current flows through the QPC for any driving frequency.

Finally, we note that if the chains are initially filled with fermions at infinite temperature (but, possibly, with different particle densities), the heating rate $\overline{\mathscr{W}}$ is zero for any driving frequency. This immediately follows from eq.~\eqref{dE}, since $\Gamma_\omega(E)$ is an odd function of $E$ and  $\rho_L$, $\rho_R$ do not depend on $E$ at infinite temperature.

\medskip
\noindent{\it Numerics.}~ Let us address numerically other types of QPCs. We start from a {\it tunneling} QPC given by
\be\label{tunneling QPC}
J_t=\sin \omega t,\quad U^L_t=U^R_t=0.
%V_t=V_t^{\rm c}=-\frac12 \sin \omega t \, (c_1^\dagger b_1+ b_1^\dagger c_1)-\frac12 \cos \omega t \, (b_1^\dagger b_1-c_1^\dagger c_1).
\ee
%\be\label{tunneling QPC}
%V_t=V_t^{\rm tun}=-\frac12  (\epsilon(t)\, c_1^\dagger b_1+ \epsilon(t)^* \,b_1^\dagger c_1),
%\ee
%where $\epsilon$ is the tunneling amplitude.
The average heating rate and current are calculated numerically and presented in Fig. \ref{Fig:main}(c) and \ref{Fig:main}(d), respectively. One can see that the phase transition for the  $\overline{\mathscr{W}}$ is there. Surprisingly, the average current also  vanishes for $\omega\geq 2$. In this respect the tunneling QPC is drastically different from the conformal QPC studied above. To visualize this difference we provide an animation of the real-time and real-space dynamics for these two QPCs  in the Supplementary Material~\cite{supplementary}. This animation shows that after initial transient dynamics the flow of fermions through the tunneling QPC halts despite the persisting large bias in particle densities.

Going further, we numerically study a range of QPCs, see Supplement Material \cite{supplementary}. We explore various
time dependencies (not necessarily harmonic) and various combinations of on-site and tunneling drivings.
We find that the heating rate $\overline{\mathscr{W}}$ vanishes for $\omega\geq 2$ for all studied QPCs. As for the average current $\overline{\mathcal{J}}$, it vanishes for some but not for all QPCs. Empirically, one necessary condition for $\overline{\mathcal{J}}$ to vanish above the critical frequency is that the tunneling rate averages to zero,
\begin{equation}\label{zero current condition}
\overline{J}\equiv \int_0^\tau J_t\, dt=0.
\end{equation}
This intuitively plausible condition is further supported by the perturbative analysis, see below. This condition is not sufficient, however, as can be seen from the case of the conformal QPC. We observe that whenever, in addition, $U^{L,R}_t$ do not depend on time,  $\overline{\mathcal{J}}$  vanishes for $\omega\geq 2$. There are some other cases when the current vanishes above the critical frequency \cite{supplementary}.  However, an exhaustive list of criteria for this effect to occur remains unknown.% This is an interesting question for further work.

\medskip
\noindent{\it Perturbative analysis.}~ For simplicity we focus on the simplest time dependence $V_t=V e^{i\omega t}+V^\dagger e^{-i\omega t}+\overline V$, which covers previously considered conformal \eqref{conformal QPC} and tunneling \eqref{tunneling QPC}  QPCs. Assuming that $V_t$ is small, we construct the Floquet Hamiltonian perturbatively in leading order, $H_{\rm F}^{(1)}=H_L+H_R+ W^{(1)}$. We calculate matrix elements of $W^{(1)}$ in the basis of eigenstates $|\zeta p\rangle$ of the Hamiltonian $H_L+H_R$ of two disconnected chains, where $p$ is the quasimomentum and $\zeta=L,R$ discriminates between the left and the right chains. The result reads \cite{supplementary}
\begin{equation}\label{W1}
W^{(1)}_{\zeta p; \eta q} = V_{\zeta p; \eta q} \frac{E_p - E_q}{E_p - E_q+\omega}+V_{\eta q;\zeta p}^* \frac{E_p - E_q}{E_p - E_q-\omega}+\overline V_{\zeta p; \eta q}.
\end{equation}
We remind the reader that $E_p=-\cos\,p$ is the energy of the disconnected chain.

In the leading order the long-time behavior of observables can be addressed via Fermi's golden rule with $W^{(1)}$ considered as the perturbation. Within this approach, the Floquet resonances at $E_p= E_q\pm \omega$ in $W^{(1)}$ are responsible for the linear growth of $\mathcal{E}$ and $N_R$ with time. Note that the first two terms in Eq. \eqref{W1} vanish for $E_p=E_q$ and therefore do not cause elastic transitions between states with the same energy.

\begin{figure}[t] %  figure placement: here, top, bottom, or page
		\centering
		\includegraphics[width=\linewidth]{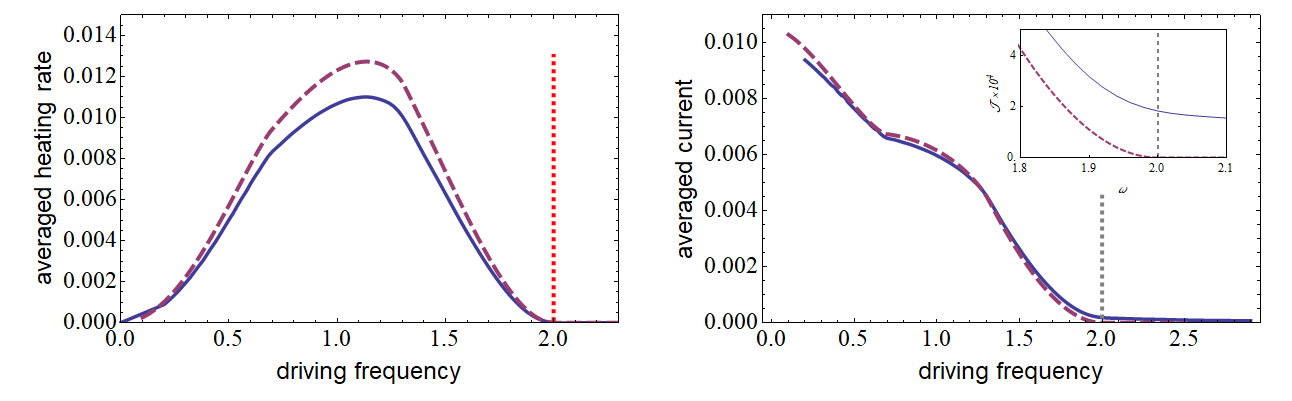}
		\caption{Averaged heating rate  $\overline{\mathscr{W}}$ (left panel) and current  $\overline{\mathcal{J}}$ (right panel) for a QPC with $J_t=0.3 \sin \omega t,\quad U^L_t=-U^R_t=  0.3\cos \omega t$ calculated numerically (solid lines, the system size and initial state the same as in Fig. \ref{Fig:main}) and perturbatively  (dashed lines, see eqs. \eqref{dE pert} and \eqref{dQ pert}, respectively). Both the average heating rate and current vanish for $\omega\geq2$ in the leading order. However, the actual current remains finite due to the higher order corrections, as illustrated in the inset to the right panel. }
		\label{Fig:pert}
\end{figure}

We find it convenient to  parameterize $J_t$, $U^{L}_t$ and $U^{R}_t$ in Eq. \eqref{VTT} as $J_t=(J e^{i\omega t}+J^* e^{-i\omega t})/2+\overline J$ and $U^{L,R}_t=(U_{L,R}\, e^{i\omega t}+U_{L,R}^* \,e^{-i\omega t})/2+\overline{U}_{L,R}$. Then we obtain in the leading order \cite{supplementary}
\begin{align}\label{dE pert}
%\overline{\mathscr{W}}^{(1)} = \int \frac{dE}{2\pi}\sum_{\eta=L,R}\left(
%1+\frac{\left|U^\eta\right|^2}{|J|^2}\right)\rho_\eta(E)T_\omega^{(1)},
%\overline{\mathscr{W}}^{(1)} = \int \frac{dE}{2\pi\tau}\sum_{\eta=L,R} \frac{|J|^2+U_\eta^2}{2|J|^2}\rho_\eta\,\Gamma_\omega^{(1)},
\overline{\mathscr{W}}^{(1)} = &\int \frac{dE}{2\pi\tau}\Bigg( \frac{|J|^2+|U_L|^2}{2|J|^2}\,\rho_L+\frac{|J|^2+|U_R|^2}{2|J|^2}\,\rho_R\Bigg)\,\Gamma_\omega^{(1)},\\[1 em]
%\end{equation}
%\begin{equation}
\label{dQ pert}
\overline{\mathcal{J}}^{(1)} = & \int \frac{dE}{2\pi} \left(\rho_L-\rho_R\right)T_\omega^{(1)}
\end{align}
with
\begin{align}
\Gamma_\omega^{(1)}=&4\pi |J|^2\,\sqrt{1-E^2} \,\, {\rm Re} \Big[\sqrt{1-(E+\omega)^2}\nonumber \\
& ~~~~~~~~~~~~~~~~~~~~~~~~~~~ - \sqrt{1- (E-\omega)^2}\Big]
\end{align}
\begin{align}
T_\omega^{(1)}=&\sqrt{1-E^2} \,\Bigg(|J|^2\,{\rm Re} \Big[\sqrt{1-(E+\omega)^2} \nonumber\\
&~~~~~~ + \sqrt{1- (E-\omega)^2}\Big]+4 | \overline J|^2 \,\sqrt{1-E^2}\Bigg)
\end{align}
where the argument $E$ is dropped in functions $\rho_{L,R}(E)$, $\Gamma_\omega^{(1)}(E)$ and $T_\omega^{(1)}(E)$ for brevity.
In Fig. \ref{Fig:pert} we compare these perturbative results with numerical calculations. One can see a good agreement even for a not-so-small~$V$.

For $\omega\geq 2$ one obtains $\Gamma_\omega^{(1)}=0$. As a consequence, the heating rate \eqref{dE pert} vanishes for any QPC in the leading order, in agreement with our numerical findings.

As for the current, it vanishes {\it in the leading order} whenever the condition \eqref{zero current condition} is satisfied. Indeed, this condition entails  $T_\omega^{(1)}=0$ for $\omega \geq 2$. It should be emphasized, however, that the condition \eqref{zero current condition} alone is not sufficient to guarantee the vanishing of the current in subsequent orders. This is illustrated in Fig. \ref{Fig:pert} and further discussed in the Supplementary Material \cite{supplementary}.

Let us also remark that if $\overline J$ is finite, the Floquet Hamiltonian above the critical frequency is given by $H_L+H_R+\overline V$ in leading order in $V_t$. The same result is straightforwardly obtained in the leading order of the Floquet-Magnus expansion \cite{bukov2015universal}. The transport through the time-independent QPC described by this Hamiltonian has been studied in detail, see \cite{Gamayun_2020,Ljubotina_2019,Doyon_2019,Lesovik_2011}.

%\cite{Landauer_1957,doi:10.1080/14786437008238472,PhysRevLett.57.1761}

\medskip
\noindent{\it Summary and discussion.}~ To summarize, we have established a nonequilibrium phase transition in a system~\eqref{Ht} of two fermionic chains filled (equally or unequally) by noninteracting fermions and connected by a periodically driven QPC. Namely, when the driving frequency~$\omega$ exceeds the bandwidth, the heating rate vanishes exactly for a generic QPC.  Furthermore, for certain QPCs the current averaged over the period also vanishes, even in the presence of a filling bias between the chains. We have verified this picture by (i) calculating the heating rate~\eqref{dE}  and the averaged current~\eqref{dQ} explicitly  for the exactly solvable conformal QPC~\eqref{conformal QPC}, (ii) performing extensive numerical studies of various QPCs and (iii) performing a perturbative analysis in the leading order in the limit when the QPC Hamiltonian can be considered as a perturbation.

It should be emphasized that vanishing of the heating rate in periodically driven systems {\it in the limit} of infinite frequency is a well-known fact that can be proven in full generality  \cite{abanin2015exponentially}. Here we obtain a much stronger result -- exact vanishing of the heating rate above a {\it finite} critical frequency.

Let us put our results in the context of prior work. First we discuss the heating rate. It is believed that generic periodically driven many-body systems (without disorder) in the thermodynamic limit heat indefinitely~\cite{dalessio2014long-time,lazarides2014equilibrium,ponte2015periodically}. On the other hand, it is commonly appreciated that dynamically integrable systems of various types can violate this rule~\cite{dalessio2014long-time,gritsev2017integrable}. For example, in the quantum Ising model with periodically driven transverse magnetic field the heating rate vanishes (after an initial transient dynamics) for any driving frequency \cite{russomanno2012periodic}. This can be shown explicitly thanks to the fact that this many-body model can be factorized into a collection of decoupled driven two-level systems~\cite{russomanno2012periodic}. More intriguingly, it has been found that the heating rate in  a kicked spin system~\cite{dalessio2013many-body}  and  a system of coupled Kapitza pendulums~\cite{citro2015dynamical}  vanishes above a critical frequency (this effect has been referred to as {\it energy localization transition}). These two systems allegedly are not dynamically integrable in any way: the vanishing of heating has been established numerically~\cite{citro2015dynamical,dalessio2013many-body} and supported by a high-frequency expansion and a variational analysis~\cite{dalessio2013many-body}. Here we establish this energy localization transition in a very different setting of a locally driven many-body system, and demonstrate its universality in this setting. We note that although we deal with noninteracting fermions, our system does not factorize into decoupled two-level systems as in Ref.  \cite{russomanno2012periodic}.

Next we discuss  transport phenomena related to our findings. The most relevant one is the phenomenon of coherent destruction of tunneling \cite{grossmann1991coherent},  where the tunneling probability through a potential barrier in a driven system vanishes at certain frequencies. This phenomenon has been established, in particular, for tight-binding lattices connected by a QPC with an oscillating local potential but constant tunneling term ($J_t={\rm const}$) \cite{reyes2017transport,agarwala2017effects}.    A related phenomenon is the real-space dynamical localization of a particle in a periodically tilted lattice~\cite{dunlap1986dynamic,kayanuma2008coherent} occurring, again, for a discrete set of frequencies. In contrast to these phenomena, the vanishing of particle flow discovered in the present work takes place for an arbitrary frequency above the critical one. Note that time dependence of the tunneling term is instrumental for this phase transition to occur.

A remark on the Floquet-Magnus expansion is in order. The Floquet-Magnus expansion is a formal expansion of the Floquet Hamiltonian in the powers of $1/\omega$~\cite{kuwahara2016floquet-magnus}. It is widely used to approximate Floquet Hamiltonians of few-level systems at high frequencies. However, its applicability to many-body systems is limited: In general, it has zero convergence radius for a generic many-body system in the thermodynamic limit \cite{blanes2009magnus,Bukov_2015,kuwahara2016floquet-magnus} (see also \cite{haga2019divergence}). In our case, the inapplicability of the Floquet-Magnus expansion can be anticipated from the fact that the exact  Floquet Hamiltonian \eqref{HFl} is {\it linear} in $\omega$. Further, it can be easily verified  \cite{supplementary} that the truncated Floquet-Magnus series contains only short-range hopping terms, while the exact Floquet Hamiltonian \eqref{HFl} contains hoppings over the entire system. As a consequence, the truncated Floquet-Magnus expansion can not be a reliable approximation of the true Floquet Hamiltonian. In particular, the nonequilibrium phase transition is not reproduced by the Floquet-Magnus expansion.   We discuss this issue in more detail in the Supplementary material~\cite{supplementary}.

We also note that one may attempt to get further analytical insight into the phenomena discussed here by means of the Floquet-Green function formalism \cite{Martinez_2003,camalet2003current,arrachea2005Green-function}. This approach remains for further work.

% Thanks to the exact solution for the conformal QPC, we are able to explicate why the true Floquet Hamiltonian \eqref{HFl} can not be fairly approximated by the Floquet-Magnus expansion truncated at a finite order $n$.

%Dynamical localization in real space \cite{dunlap1986dynamic,kayanuma2008coherent} for a discrete set of frequencies (experimental verification in \cite{lignier2007dynamical,eckardt2009exploring}). Destroyed by adding interactions \cite{luitz2017absence}.

Finally, we briefly remark on possible ways to test our predictions experimentally. A well-developed quantum dot technology  provides a necessary toolbox for this task \cite{Kaestner_2015}. Another, recently emerged option is to use a cold atom simulator of a quantum point contact \cite{krinner2015observation,Krinner_2017,Lebrat2018band}. The latter platform benefits from the perfect isolation from the environment and  extended control over the effective Hamiltonian.  According to our findings,  a QPC can act as a frequency-controlled quantum switch, and experimental observations of this effect may pave the way to its technological applications.

%Solid-state driven quantum dots are well developed  of the present work can be tested with the  experiments with  quantum dots, cold atoms or  trapped ions. Furthermore, our theoretical results show that the QPC can act as a frequency-controlled quantum switch, which may be useful in technological applications.

\begin{acknowledgments}
\textit{Acknowledgments.}
O. G. and J.-S.C. acknowledge the support from the European Research Council under ERC Advanced grant 743032 DYNAMINT. The work of O.L. (numerical simulations and contributions to the perturbative analysis) is supported by
the Russian Science Foundation under the grant No 17-12-01587.
\end{acknowledgments}

%\bibliography{confDlitra,Floquet,dynamically_integrable,LZ_and_adiabaticity}
\bibliography{C:/D/Work/QM/Bibs/confDlitra,C:/D/Work/QM/Bibs/Floquet,C:/D/Work/QM/Bibs/dynamically_integrable,C:/D/Work/QM/Bibs/LZ_and_adiabaticity}
\bibliographystyle{apsrev4-1}

\clearpage

%\newpage

%\appendix

\renewcommand{\theequation}{S\arabic{equation}}
\setcounter{equation}{0}

\renewcommand{\thefigure}{S\arabic{figure}}
\setcounter{figure}{0}

\onecolumngrid

%\begin{widetext}

\section{{\large Supplementary material}}

%\begin{widetext}

\section{S1. Spectral properties}

\begin{figure}[h] %  figure placement: here, top, bottom, or page
	\centering
	\includegraphics[width=\textwidth]{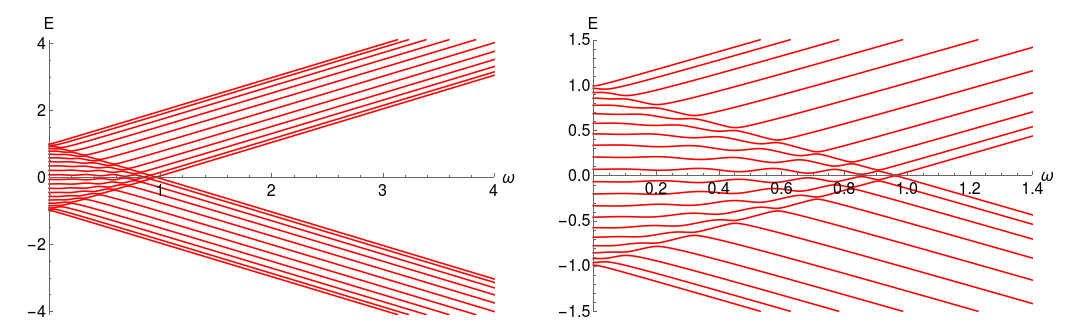}
	\caption{The single-particle spectrum of the Floquet Hamiltonian $H^\omega$ with $L=11$  for different driving frequencies.}
	\label{Fig:spec}
\end{figure}

In this section we describe spectral properties of the Floquet Hamiltonian (7) of the conformal defect. At the single-particle level this Hamiltonian is represented by the $2L\times 2L$ matrix
\begin{equation}
H^\omega = H_0 +\frac{\omega}2\, \Sigma\label{Hw}
\end{equation}
with
\begin{equation}
H_0 =\left( \begin{array}{cc}
H_{0L} & 0 \\
0 & H_{0R}
\end{array}\right),\qquad  \Sigma=-i \left(\begin{array}{cc}
0 & -\sigma \\
\sigma & 0
\end{array}\right). \label{H0}
\end{equation}
Here $H_{0L}$ ($H_{0R}$) describes respectively the left (right) tight-binding chain, {\it including} the initial on-site potentials at sites $L$ and $L+1$ (not to be confused with $H_L$ and $H_R$ introduced in the main text). In the lattice sites basis
these matrices are given by
\begin{equation}
H_{0L}=-\frac12\left(\begin{array}{cccccc}
0 & 1  & 0  & 0 & \dots & 0 \\
1  & 0 & 1  & 0 & \dots & 0 \\
0  & 1  & 0 & 1 & \dots & 0 \\
\vdots  & \vdots &\vdots & \ddots & \vdots & \vdots \\
0  & \dots  & 0 & 1 & 0 & 1 \\
0  & \dots  & 0 & 0 & 1 & 1
\end{array}\right),\quad
H_{0R}=-\frac12\left(\begin{array}{cccccc}
-1 & 1  & 0  & 0 & \dots & 0 \\
1  & 0 & 1  & 0 & \dots & 0 \\
0  & 1  & 0 & 1 & \dots & 0 \\
\vdots  & \vdots &\vdots & \ddots & \vdots & \vdots \\
0  & \dots  & 0 & 1 & 0 & 1 \\
0  & \dots  & 0 & 0 & 1 & 0
\end{array}\right),\quad  \sigma=\left(\begin{array}{ccccc}
0 & 0 &\ldots  & 0 & 1  \\
0 & 0 & \ldots  & 1  &  0\\
\vdots & \vdots & \ddots & \vdots &\vdots \\
0 & 1 & \ldots & 0 & 0 \\
1 & 0 & 0 & 0 & 0
\end{array}\right).
\end{equation}
The diagonalization can be achieved by the appropriate Fourier transformation. To present the spectrum and eigenvectors
in concise form we introduce auxiliary notations. First we parametrize energy $E$ and the driving frequency $\omega$
with parameters $z_\pm$
\begin{equation}
E = - \frac{\cosh(z_+)+\cosh(z_-)}{2},\qquad \omega = -\cosh(z_+)+\cosh(z_-)
\end{equation}
Conversely,
\begin{equation}\label{zpmDef}
-\cosh(z_\pm)=E\pm \omega/2.
\end{equation}
The spectrum can be found as zeroes of the function
\begin{equation}\label{gDef}
g(E) =w(E)-u(E),
\end{equation}
where
\begin{equation}\label{wDef}
w(E) =  \frac{s_+(L)}{s_+(L+1)},\qquad u(E) =  \frac{s_-(L+1)}{s_-(L)},\qquad
s_{\pm}(x) \equiv\sinh(z_\pm x).
\end{equation}
The eigenvectors of this Hamiltonian can be represented as
\begin{equation}\label{E}
|E\rangle=\frac{1}{\sqrt{g'(E)}}\sum_{k=1}^L\left(\frac{s_+(k)}{s_+(L+1)}+\frac{s_-(k)}{s_-(L)}\right)c_k^\dag|0\rangle  +
\frac{i}{\sqrt{g'(E)}}\sum_{k=L+1}^{2L} \left(\frac{s_+(2L+1-k)}{s_+(L+1)}-\frac{s_-(2L+1-k)}{s_-(L)}\right)c_k^\dag|0\rangle.
\end{equation}
Notice that the coefficients are nothing but a ratio of Chebyshev polynomials of the second kind of different degrees that depend either on $-E-\omega/2$
or $-E+\omega/2$, which renders them real for real $E$. Here he have chosen the normalization such that $g'(E)$ is positive in all spectral points.
The typical form of the spectrum with respect to the driving frequency is shown in Fig. \eqref{Fig:spec}.

The spectral data has a smooth limit to the non-driven case. Indeed, in the $\omega \to 0$ limit we put $z_-=z_+ =z$, which transforms the spectrum condition into
\begin{equation}
g(E) = \frac{\cosh (2Lz)-\cosh(2(L+1)z)}{2\sinh(Lz)\sinh((L+1)z)}=0,\qquad E = -\cosh(z).
\end{equation}
This condition can be easily resolved and the spectrum is given by
\begin{equation}\label{spe0}
2(L+1)z  =- 2Lz +2\pi  i q ~~\Rightarrow~~ z = \frac{\pi i q}{2L+1} \equiv i\varphi_q~~\Rightarrow~~ \mathrm{E}_q = - \cos \varphi_q,\quad q=1,2,\dots,2L.
\end{equation}
The corresponding eigenvector reads
\begin{equation}
|\Phi_q\rangle=\left[\sum_{k=1}^L \frac{1-(-1)^{q}}{\sqrt{2L+1}}\sin\Big(\frac{qk\pi}{2L+1}\Big) c_k^\dag|0\rangle
+i\sum_{k=L+1}^{2L} \frac{1+(-1)^{q}}{\sqrt{2L+1}}\sin\Big(\frac{qk\pi}{2L+1}\Big)c_k^\dag|0\rangle\right].
\end{equation}

In the thermodynamic limit we characterize the spectrum by the density of states $D(E)$. It is defined such that given a continuous function $f(E)$ the following sum has a limiting integral form:
\begin{equation}\label{test1}
\lim\limits_{L\to \infty} \frac{1}{2L} \sum\limits_{n=1}^{2L}	 f(E_n) =  \int \frac{dE}{2\pi} D(E )f(E).
\end{equation}
In this case it is more convenient to change the spectrum function as
\begin{equation}
g(E) \to g(E) = \frac{\sinh(z_+(L+1))}{\sinh z_+} \frac{\sinh(z_-(L+1))}{\sinh z_-} - \frac{\sinh(z_+ L)}{\sinh z_+} \frac{\sinh(z_-L)}{\sinh z_-}
\end{equation}
Then we can represent the sum in Eq. \eqref{test1} as the contour integral
\begin{equation}
\sum\limits_{n=1}^{2L}f(E_n)  = \frac{1}{2L} \oint_\gamma \frac{dE}{2\pi i} \frac{g'(E)}{g(E)}f(E),
\end{equation}
where the contour $\gamma$ encircles only the poles that come from $g(E)$.
In this form we can take the $L \to \infty$ limit to obtain
\begin{equation}
D(E) = \frac{1}{2} {\rm Re} \left(\frac{1}{\sqrt{1-(E-\omega/2)^2}}+\frac{1}{\sqrt{1-(E+\omega/2)^2}}\right).
\end{equation}
This formula is valid for all $\omega$ and, in particular, for $\omega = 0 $ it reproduces the density of state for the spectrum \eqref{spe0}.
Notice that for $0<\omega <2$ the density of states contains integrable singularities not only at the spectral boundaries but also inside the spectrum.

\section{S2. Current and heating rate}

We consider the initial wave function to be a (possibly, mixed state) which is stationary with respect to  $H_{t=0}$. In other words, initially each one-particle state $|\Phi_q\rangle$ is filled with some probability $\rho_q$.\footnote{Note that $H_{t=0}$ is different from $H_L+H_R$ since it contains on-site potentials at two adjacent edges of the chain. Accordingly, the initial state of the system is locally different from an equilibrium state of two disjoint chains without on-site potentials. However, this local difference of the initial condition does not affect the steady state in the long-time limit.}
Our  observables of interest are the current through the defect and the energy pumped into the system over a full period.
To compute the current we consider the expectation value of the number $N_R$ of particles in the right part of the system:
\begin{equation}
\langle N_R(t) \rangle = \sum_q \rho_q\langle \Phi_q|N_R(t)|\Phi_{q}\rangle.
\end{equation}
The time dynamics can be taken into account by the form-factor expansion
\begin{align}
\langle \Phi_q|N_R(t)|\Phi_{q}\rangle=&\sum_{p,p'}\langle \Phi_q|E_p\rangle\langle E_p| e^{iH^\omega t}e^{-i\omega t \Sigma /2}N_R e^{i\omega t \Sigma /2}e^{-iH^\omega t}|E_{p'}\rangle\langle E_{p'}|\Phi_{q}\rangle=\nonumber \\
=&\sum_{p,p'}e^{i(E_p-E_{p'})t}\langle\Phi_q|E_p\rangle\langle E_p|e^{-i\omega t \Sigma /2} N_Re^{i\omega t \Sigma /2}|E_{p'}\rangle\langle E_{p'}|\Phi_{q}\rangle
\end{align}
The matrix elements are as follows
\begin{equation}
\langle E_p|e^{-i\omega t \Sigma /2} N_Re^{i\omega t \Sigma /2}|E_{p'}\rangle=\frac{\delta_{pp'}}{2} - \frac{1-w_pw_{p'}}{2\sqrt{g'(E_p)g'(E_{p'})}} \left(
\frac{e^{-i\omega t}}{E_p-E_{p'}-\omega} + \frac{e^{i\omega t}}{E_{p'}-E_{p}-\omega}
\right).
\end{equation}
Here we introduce the notation $w(E_p)\equiv w_p$ for the function defined in Eq. \eqref{wDef}.
For the current computation we need the time derivative which can be represented as
\begin{equation}
\frac{d}{dt} e^{it(E_p-E_{p'})}\langle E_p|U_{-\epsilon(t)} N_RU_{\epsilon(t)}|E_{p'}\rangle = -\sin(\omega t)e^{it(E_p-E_{p'})}\frac{1-w_pw_{p'}}{\sqrt{g'(E_p)g'(E_{p'})}}.
\end{equation}
Similarly, the total energy in the system at time $t_n = 2\pi n/\omega$, where $U(t_n)=\pm 1$, is given by the expansion
\begin{equation}
\mathcal{E}_n = 	\sum_q \rho_q\langle \Phi_q |e^{i t_n H^\omega}H_0e^{-i t_n H^\omega}| \Phi_q \rangle  =
\sum_q \rho_q \langle \Phi_q |E_p \rangle \langle E_p | H_0| E_{p'}\rangle \langle E_{p'}| \Phi_q \rangle e^{it_n(E_p-E_{p'})}.
\end{equation}
Taking into account that
\begin{equation}
(H_0+\frac{1}2\omega \, \Sigma) |E_p\rangle = E_p|E_p\rangle,
\end{equation}
the matrix element of $H_0$ reads
\begin{equation}
\langle E_{p}|H_0 | E_{p'} \rangle = E_{p} \delta_{pp'} - \frac{1}2\omega
\langle E_p | \Sigma |  E_{p'}\rangle
\end{equation}
and moreover, the matrix elements of the $\Sigma$ can be easily evaluated
\begin{equation}
\langle E_p | \Sigma |  E_{p'}\rangle  \sqrt{g'(E_p)g'(E_{p'})} =-\frac{2(w_p-w_{p'})}{E_p-E_{p'}}.
\end{equation}
This way, the amount of energy pumped into the system over a period can be written as
\begin{equation}
\Delta \mathcal{E}_n \equiv  \mathcal{E}_{n+1} - \mathcal{E}_n = i \omega  \int\limits_{t_n}^{t_{n+1}} dt
\rho_q\sum_q \langle \Phi_q |E_p \rangle \langle E_{p'}| \Phi_q \rangle (w_p-w_{p'})e^{it(E_p-E_{p'})}.
\end{equation}
The last ingredient we need is the overlap between $|\Phi_q\rangle$ and an eigenstate that
corresponds to the finite frequency $\omega$  \eqref{E}. To compute it we notice that
for odd (even) $q$ only the left (right) part of the chain is filled in accordance with the block structure of the operator \eqref{H0}.
Therefore, the global phase can be chosen separately for the left and right parts. This way, the overlap reads as
\begin{equation}
\langle \Phi_q|E_p\rangle = \frac{\sin(\varphi_q(L+1))}{\sqrt{2L+1}\sqrt{g'(E_p)}} \frac{\omega (1+(-1)^q w_p)}{(E_p-\mathrm{E}_q)^2-(\omega/2)^2}.
\end{equation}
Observe the difference between the eigenenergy $E_p$ of $H^\omega$ and eigenenergy $\mathrm{E}_p$ of  $H_0$.\footnote{Also note that $E_p$ has a different meaning in the main text and in Sect. S5.}
Taking into account
\begin{equation}
1-w_pw_{p'} = \frac{(1-w_p)(1+w_{p'})}{2} + \frac{(1+w_p)(1-w_{p'})}{2},\qquad w_p-w_{p'} = -\frac{(1-w_p)(1+w_{p'})}{2}+\frac{(1+w_p)(1-w_{p'})}{2}
\end{equation}
and introducing functions
\begin{equation}
\mathcal{A}_{\pm}(q,t) =\sum_p \frac{e^{iE_pt}}{g'(E_p)} \frac{\omega (1+(-1)^q w_p)(1\pm w_p)}{(E_p-\mathrm{E}_q)^2-(\omega/2)^2}\label{AA}
\end{equation}
we obtain
\begin{equation}
J_R = \frac{d \langle N_R(t) \rangle }{dt} = -\sin(\omega t) \sum_q \rho_q\frac{\sin^2(\varphi_q(L+1))}{2L+1}{\rm Re}\mathcal{A}_-(q,t) \mathcal{A}^*_+(q,t),
\end{equation}
\begin{equation}
\Delta \mathcal{E}_n =\omega  \int\limits_{t_n}^{t_{n+1}} dt \sum_q\rho_q\frac{\sin^2(\varphi_q(L+1))}{2L+1}{\rm Im}\mathcal{A}_-(q,t) \mathcal{A}^*_+(q,t).
\end{equation}
To effectively evaluate these functions we notice that in the spectral points $w_p = u_p\equiv u(E_p)$, (see Eqs. \eqref{wDef}), therefore, we can rewrite
functions $\mathcal{A}_\pm$, (Eqs. \eqref{AA}), to avoid singularities at $E_p =\mathrm{E}_q \pm \omega/2$. Another simplification can be achieved by rendering the numerators
as linear functions in $u(E)$  and $w(E)$, which is needed to perform summation by contour integral and is motivated by the fact that the spectral function $g(E)$ is also a linear combintation of $u(E)$ and $v(E)$. With all these requirements we obtain
\begin{equation}
\mathcal{A}_-(q,t) = \sum_p \frac{e^{iE_pt}}{g'(E_p)}  \left(
\frac{(1+(-1)^q u_p)(1-w_p)}{E_p-\mathrm{E}_q-\omega/2}-\frac{(1+(-1)^q w_p)(1-u_p)}{E_p-\mathrm{E}_q+\omega/2}
\right),\label{A}
\end{equation}
\begin{equation}
\mathcal{A}_+(q,t) = \sum_p \frac{e^{iE_pt}}{g'(E_p)}  \left(
\frac{(1+(-1)^q u_p)(1+w_p)}{E_p-\mathrm{E}_q-\omega/2}-\frac{(1+(-1)^q w_p)(1+u_p)}{E_p-\mathrm{E}_q+\omega/2}
\right).\label{B}
\end{equation}
We compute these functions exactly  in the thermodynamic limit in the next section of this Supplementary Material (see Eqs. \eqref{al}-\eqref{br}).
For odd and even $q$ they can be presented as regular functions $\mathcal{A}^l_{\pm}(\mathrm{E}_q,t)$, and $\mathcal{A}^r_{\pm}(\mathrm{E}_q,t)$ respectively.
Moreover, one can notice the following reflection relations
\begin{equation}
\mathcal{A}^r_-(E,t) = - [\mathcal{A}_+^l(-E,t)]^*,\qquad \mathcal{A}^l_-(E,t) = [\mathcal{A}_+^r](-E,t)^*,
\end{equation}
which leads to
\begin{equation}
\mathcal{A}^l_-(E,t) [\mathcal{A}^l_+(E,t)]^*  = - \mathcal{A}^r_-(-E,t) [\mathcal{A}^r_+(-E,t)]^*\equiv \frac{\sqrt{1-E^2}}{1-E}\mathcal{F}(E,t).
\end{equation}
Here the notation $\mathcal{F}(E,t)$ is introduced accounting for the density of states in the non-driven case and the fact that $\sin^2(\varphi_q(L+1))= (1+(-1)^q\mathrm{E}_q)/2$. Therefore, if we characterize the initial state in which the odd quantum numbers are filled with the density $\rho_L(E)$ and the even with the density $\rho_R(E)$, the current can be represented as
\begin{equation}
J_R = -\sin(\omega t) \int dE \frac{\rho_L(E)-\rho_R(-E)}{4\pi} {\rm Re} \mathcal{F}(E,t).
\end{equation}
Using large time asymptotic of the functions $\mathcal{A}^l_\pm(E,t)$ one can conclude the following time-dependence
\begin{equation}\label{F}
\mathcal{F}(E,t) =
F(E) + e^{i \omega t} F_+(E) + e^{-i \omega t} F_-(E)
\end{equation}
therefore after averaging over a period, the current takes a Landauer-B\"uttiker like expression with the effective transmission coefficient $T(E)$
\begin{equation}\label{J}
\bar{J}_R =  \int \frac{dE}{2\pi} (\rho_L(E)-\rho_R(E))T(E)
\end{equation}
with $ T(E) = \frac{1}{4} {\rm Im} \left[F_+(E) - F_-(E)\right]$.
Similarly, for the energy pumped through the period
\begin{equation}\label{dE suppl}
\Delta \mathcal{E} = \int \frac{dE}{2\pi}(\rho_L(E)+\rho_R(E))\Gamma(E),
\end{equation}
with the heating function $\Gamma(E) =\pi {\rm Im} F(E)$.
Here we have used that $T(E)=T(-E)$  and $\Gamma(E)=-\Gamma(-E)$.

The explicit form of the transmission coefficient reads
\begin{multline}
T(E) ={\rm Re}\left[ (1-E^2) \left(1-\left(\frac{\sqrt{(E-\omega)^2-1}+\sqrt{(E+\omega)^2-1}}{2\omega}\right)^2\right) +\right.\\ \left.
\frac{\sqrt{1-E^2}}{2\omega^2} \left(\sqrt{1-(E-\omega)^2}(E^2+E\omega-1)+\sqrt{1-(E+\omega)^2}(E^2-E\omega-1)\right)\right]
\end{multline}
%  \begin{figure}[h] %  figure placement: here, top, bottom, or page
% 	\centering
% 	\includegraphics[width=0.9\textwidth]{trans.png}
% 	\caption{The effective transmission coefficient for different driving frequencies}
% 	\label{trans}
% \end{figure}
% The typical behavior is shown in Fig. \eqref{Fig:se}.
Notice the non-smooth behavior for frequencies $0<\omega/2<1$.
Low and high frequency expansions are as follows
\begin{equation}
T(E) \overset{\omega\to 0}{=} \frac{1}{2},\qquad
T(E) \overset{\omega\to \infty}{=} \frac{1-E^2}{\omega^2}.
\end{equation}
Similarly the heating function $\Gamma(E)$ reads
\begin{multline}
\Gamma(E) =2\pi\frac{\sqrt{1-E^2}}{\omega^2} {\rm Re} \left[(\sqrt{1-(E-\omega)^2}(E^2 - E\omega-1)- \sqrt{1- (E+\omega)^2}(E^2 + E\omega-1))\right]+ \\
+ 2\pi\frac{1-E^2}{\omega^2} \left(
((E+\omega)^2-1) \theta(1- \omega -E) -
((E-\omega)^2-1) \theta(E- \omega +1)
\right).
\end{multline}
Here $\theta(\dots)$ are the step functions. The most remarkable feature of the heating function is that it is identically zero for high frequencies
\begin{equation}
\Gamma(E) =0 ,\qquad \omega >2.
\end{equation}

\section{S3. Summation procedures}

In this appendix we describe the way to evaluate sums in Eqs. \eqref{A}, \eqref{B}.
We provide detailed computation for the first sum of Eq. \eqref{A} for odd $q$ while for others we just give final results.
First we transform the sum in interest as the contour integral
\begin{equation}
S= \sum_p \frac{e^{iE_pt}}{g'(E_p)}
\frac{(1- u_p)(1-w_p)}{E_p-\mathrm{E}_q-\omega/2} = \oint\limits_\gamma \frac{dE}{2\pi i} \frac{e^{iEt}}{g(E)} \frac{(1- u(E))(1-w(E))}{E-\mathrm{E}_q-\omega/2}
\end{equation}
where the contour $\gamma$ goes in the counterclockwise direction around the spectrum points. The numerator has been chosen such that there is no singularity at $E = \mathrm{E}_q+\omega/2$,
therefore, instead of each individual point we encircle
the union of the intervals $I = I_- \cup I_+ $ defined as
$I_\pm = [-1\pm \omega/2, 1\pm \omega/2]$ (here it is assumed that $\omega>0$). This way we obtain
\begin{equation}
S= \int_I \frac{e^{iEt} dE}{2\pi i} \left(\frac{(1- u(E))(1-w(E))}{g(E)(E-\mathrm{E}_q-\omega/2)}\Big|_{E\to E-i0}-
\frac{(1- u(E))(1-w(E))}{g(E)(E-\mathrm{E}_q-\omega/2)}\Big|_{E\to E+i0}\right).
\end{equation}
In this form one can immediately take the thermodynamic limit $L \to \infty$. The limiting form of the functions $w(E\pm i0)$ and $u(E\pm i0)$ would be determined
by those solution $z_\pm$ of Eqs. \eqref{zpmDef} with positive imiginary parts. To facilitate its finding we notice that the general solution of the equation
\begin{equation}
\cosh y = E + i\eta 0,
\end{equation}
with the positive real part can be described as follows
\begin{equation}
e^{y} = \left\{\begin{array}{ll}
E+\sqrt{E^2-1}, & E>1\\
E-\sqrt{E^2-1}, & E<-1\\
E+i {\rm sgn \eta}\sqrt{1-E^2}, & |E|<1
\end{array}\right.
\end{equation}
where the square root takes its positive value.
This way, we obtain
\begin{equation}
-u(E+i\eta 0) = \left\{\begin{array}{ll}
E-\omega/2-\sqrt{(E-\omega/2)^2-1}, & E \in I_-\setminus  \left(I_-\cap I_+\right)\\
E-\omega/2+i {\rm sgn}(\eta)\sqrt{1-(E-\omega/2)^2}, & E \in I_-\cap I_+\\
E-\omega/2+i {\rm sgn}(\eta)\sqrt{1-(E-\omega/2)^2}, & E \in I_+\setminus  \left(I_-\cap I_+\right)
\end{array}\right.
\end{equation}
\begin{equation}
-w(E+i\eta 0) = \left\{\begin{array}{ll}
E+\omega/2-i{\rm sgn}(\eta)\sqrt{1-(E+\omega/2)^2}, & E \in I_-\setminus  \left(I_-\cap I_+\right)\\
E+\omega/2-i{\rm sgn}(\eta)\sqrt{1-(E+\omega/2)^2}, & E \in I_-\cap I_+\\
E+\omega/2-\sqrt{(E+\omega/2)^2-1}, & E \in I_+\setminus  \left(I_-\cap I_+\right)
\end{array}\right.
\end{equation}
It is useful to introduce the following functions
\begin{equation}\label{uw}
-u_{\pm}(E) = E-\omega/2 \pm i \sqrt{1-(E-\omega/2)^2},\qquad
-w_{\pm}(E) = E+\omega/2 \mp i\sqrt{1-(E+\omega/2)^2}
\end{equation}
Then assuming that $\sqrt{-1} = +i$, we can present
the above cases as
\begin{equation}\label{uu}
u(E\pm i 0) = \left\{\begin{array}{ll}
u_+(E), & E \in I_-\setminus  \left(I_-\cap I_+\right)\\
u_\pm(E), & E \in I_+
\end{array}\right.
\end{equation}
\begin{equation}\label{ww}
w(E\pm i 0) = \left\{\begin{array}{ll}
w_\pm(E), & E \in I_-\\
w_-(E), & E \in I_+\setminus  \left(I_-\cap I_+\right)
\end{array}\right.
\end{equation}
This way the sum $S$ for $\omega>2$ can be presented as
\begin{multline}
S = \int_{I_+} \frac{e^{iEt}dE}{2\pi i} \left(
\frac{(1-u_-)(1-w_-)}{(E-\mathrm{E}_q-\omega/2-i0)(w_--u_-)} -
\frac{(1-u_+)(1-w_-)}{(E-\mathrm{E}_q-\omega/2+i0)(w_--u_+)}
\right) +  \\
\int_{I_-} \frac{e^{iEt}dE}{2\pi i} \left(
\frac{(1-u_+)(1-w_-)}{(E-\mathrm{E}_q-\omega/2-i0)(w_--u_+)} -
\frac{(1-u_+)(1-w_+)}{(E-\mathrm{E}_q-\omega/2+i0)(w_+-u_+)}
\right).
\end{multline}
Here, for brevity, we dropped $E$-dependence in functions $u_{\pm}(E)$ and $w_{\pm}(E)$.
For $0<\omega<2$ the answers can be obtained in the same manner. Further simplifications can be achieved employing the exact form of the integrands given by
Eq. \eqref{uw} and using the Sokhotski–-Plemelj formula to extract singularities.

Proceeding in the same way, we can obtain similar expression for sums in Eqs. \eqref{A}, \eqref{B}, which transforms into
regular functions of energy, that are being evaluated at the point $\mathrm{E}_q$. The corresponding limiting functions that correspond
to the odd $q$ we denote as $\mathcal{A}^l_\pm(E,t)$ and as $\mathcal{A}^r_\pm(E,t)$  for even $q$.
To write them down in a compact form we introduce the following notations
\begin{equation}
M(E) = \frac{\omega}{(E-\mathrm{E}_{q} )^2-(\omega/2)^2}
,\qquad \mathrm{E}_{q}^{\pm} = \mathrm{E}_{q} \pm \omega/2,
\end{equation}
\begin{equation}
Y_\omega (E) = (1+E+\omega/2) \frac{\sqrt{1-(E-\omega/2)^2}}{\omega},\qquad
X(E) = \frac{\sqrt{(1-(E-\omega/2)^2)(1-(E+\omega/2)^2)}}{\omega}.
\end{equation}
Then the functions from Eqs. \eqref{A},\eqref{B} read
\begin{equation}
\mathcal{A}_-^l(\mathrm{E}_q,t) = \int\limits_{I_+} \frac{dE}{\pi} e^{iEt} Y_\omega(E) M(E) + \int\limits_{I_-} \frac{dE}{\pi} e^{iEt} Y_{-\omega}(E) M(E)
 -e^{i\mathrm{E}_q^+ t }{\rm Im} Y_{-\omega} (\mathrm{E}_q^+ ) - e^{i\mathrm{E}_q^- t }{\rm Im} Y_\omega (\mathrm{E}_q^- ) \label{al}
\end{equation}
\begin{multline}
\mathcal{A}_-^r(\mathrm{E}_q,t) = \int\limits_{I_+} \frac{dE}{\pi} e^{iEt} X(E) M(E) - \int\limits_{I_-} \frac{dE}{\pi} e^{iEt} X(E) M(E)
\\  + e^{i\mathrm{E}_q^+t}\left(\frac{(\mathrm{E}_q^+)^2-(\omega/2+1)^2}{\omega}+{\rm Re}X(\mathrm{E}_q^+ ) \right)-e^{i\mathrm{E}_q^- t }\left(\frac{(\mathrm{E}_q^-)^2-(\omega/2-1)^2}{\omega}+{\rm Re}X(\mathrm{E}_q^- ) \right)\label{ar}
\end{multline}
\begin{multline}\label{bl}
\mathcal{A}_+^l(\mathrm{E}_q,t) = \int\limits_{I_+} \frac{dE}{\pi} e^{iEt} X(E) M(E) - \int\limits_{I_-} \frac{dE}{\pi} e^{iEt} X(E) M(E)
\\  + e^{i\mathrm{E}_q^+t}\left(\frac{(\mathrm{E}_q^+)^2-(\omega/2-1)^2}{\omega}+{\rm Re}X(\mathrm{E}_q^+ ) \right)- e^{i\mathrm{E}_q^- t }\left(\frac{(\mathrm{E}_q^-)^2-(\omega/2+1)^2}{\omega}+{\rm Re}X(\mathrm{E}_q^- ) \right)
\end{multline}
\begin{equation}\label{br}
\mathcal{A}_+^r(\mathrm{E}_q,t) = \int\limits_{I_+} \frac{dE}{\pi} e^{iEt} Y_\omega(-E) M(E) + \int\limits_{I_-} \frac{dE}{\pi} e^{iEt} Y_{-\omega}(-E) M(E)
- e^{i\mathrm{E}_q^+ t }{\rm Im} Y_\omega (-\mathrm{E}_q^+ ) - e^{i\mathrm{E}_q^- t }{\rm Im} Y_{-\omega} (-\mathrm{E}_q^- )
\end{equation}
All integrals above are understood in the principal value sense.
The large time asymptotic can be easily deduced using the fact that for any regular function $f(E)$
\begin{equation}
\frac{1}{\pi i}~ P\!\int\limits \frac{f(E)}{E-E_0} e^{iEt} \approx f(E_0)e^{it E_0},\qquad t\to+\infty,
\end{equation}
we obtain
\begin{equation}
\mathcal{A}^l_-(\mathrm{E}_q,t) = i\left(Y_\omega(\mathrm{E}_q^+)+Y_{-\omega}(\mathrm{E}_q^+)\right) e^{i\mathrm{E}_q^+ t }
-i\left(Y^*_\omega(\mathrm{E}_q^-)+Y_{-\omega}(\mathrm{E}_q^-)\right) e^{i\mathrm{E}_q^- t }
\end{equation}

\begin{equation}
\mathcal{A}^r_-(\mathrm{E}_q,t) = e^{i \mathrm{E}_q^+ t} \left(
X(\mathrm{E}_q^+) + \frac{(\mathrm{E}_q^+)^2-(\omega/2+1)^2}{\omega}
\right) - e^{i \mathrm{E}_q^- t} \left(
X^*(\mathrm{E}_q^-) + \frac{(\mathrm{E}_q^-)^2-(\omega/2-1)^2}{\omega}
\right)
\end{equation}

\begin{equation}
\mathcal{A}^l_+(\mathrm{E}_q,t) = e^{i \mathrm{E}_q^+ t} \left(
X(\mathrm{E}_q^+) + \frac{(\mathrm{E}_q^+)^2-(\omega/2-1)^2}{\omega}
\right) - e^{i \mathrm{E}_q^- t} \left(
X^*(\mathrm{E}_q^-) + \frac{(\mathrm{E}_q^-)^2-(\omega/2+1)^2}{\omega}
\right)
\end{equation}
\begin{equation}
\mathcal{A}^r_+(\mathrm{E}_q,t) = i\left(Y_\omega(-\mathrm{E}_q^+)+Y_{-\omega}(-\mathrm{E}_q^+)\right) e^{i\mathrm{E}_q^+ t }
-i\left(Y^*_{-\omega}(-\mathrm{E}_q^-)+Y_\omega(-\mathrm{E}_q^-)\right) e^{i\mathrm{E}_q^- t }
\end{equation}

\section{S5. Perturbative analysis\label{sec:pert}}
\subsection{S5.1. Perturbative expansion for the Floquet Hamiltonian}

Let us consider a generic time-dependent periodic perturbation
\begin{equation}
H = H_0 + \sum_s V^{(s)} e^{i\omega s t } = H_0 + V(t).
\end{equation}
The evolution operator reads
\begin{equation}
U(t) = e^{-itH_0}S(t)
\end{equation}
for $S(t)$ given by the time-ordered exponential
\begin{equation}\label{St1}
S(t) = \sum\limits_{k=1}^\infty (-i)^k \int\limits^t_0 dt_1 \int\limits^{t_1}_0 dt_2 \int\limits^{t_2}_0 \dots \int\limits_0^{t_{k-1}}dt_n V(t_1) V(t_2) \dots V(t_k).
\end{equation}
Our plan is to consider stroboscopic evolution, i.e. focusing just on times $t_n = 2\pi n /\omega$ and present it as the evolution with time independent
effective Hamiltonian $H = H_0 + W$, namely
\begin{equation}
e^{-it_n H_0}S(t_n) =  e^{-it_n(H_0+W)}.
\end{equation}
To accomplish this task we rewrite series \eqref{St1} as an integral equation
\begin{equation}\label{S(t)}
S(t)  =1 - i \int\limits^t_0 d\tau V(\tau) S(\tau).
\end{equation}

We consider matrix elements in the basis of eigenvectors of $H_L+H_R$
\begin{equation}
(H_L+H_R) |\eta,k\rangle = E_k |\eta,k \rangle,
\end{equation}
where $k\in [0,\pi]$ is a quasimomentum quantized in integers of $\pi/(L+1)$ and $\eta=L,R$ discriminates between the two chains.\footnote{ Note that the meaning of $E_k$ here is different from that in Secs. (S1)-(S4).} Explicitly,
\begin{equation}\label{eigenstates}
|L,k\rangle=\sqrt{\frac2L} \sum_{j=1}^L \sin \left(k(L-j+1)\right)\, c^\dagger_j |0\rangle,\quad |R,k\rangle=\sqrt{\frac2L} \sum_{j=L+1}^{2L} \sin \left(k(j-L)\right)\, c^\dagger_j |0\rangle,\quad E_k=-\cos k.
\end{equation}
We will often skip the index $\eta$ for brevity when this does not lead to confusion. For example, we will write down $V_{pq}$ instead of  $V_{\zeta p;\eta q}$.

Eq. \eqref{S(t)} entails
\begin{equation}
S_{pq}(t) = \delta_{pq} - i \int\limits^t_0 d\tau e^{i\tau E_{pk}}V_{pk}S_{pq}(\tau),\qquad E_{pk} = E_p - E_k.
\end{equation}
It is useful to rewrite this equation after Laplace transformation
\begin{equation}
S_{pq}(t) \to S_{pq}(\lambda ) \equiv \int\limits_0^\infty dt e^{-\lambda t}S_{pq}(t)
\end{equation}
We have
\begin{equation}
 S_{pq}(\lambda ) = \frac{\delta_{pq}}{\lambda}  - \frac{i}{\lambda} \sum_{k,s} V^{s}_{pk} S_{kq}(\lambda - iE_{pk}-i\omega s),
\end{equation}
where $E_{pq}\equiv E_p-\mathrm{E}_q$.
This equation can be formally solved by iterations. The leading terms reads as
\begin{equation}\label{flfl}
S^{(0)}_{pq} = \frac{\delta_{pq}}{\lambda} ,\qquad S^{(1)}_{pq} = \frac{-i}{\lambda} \sum_{s}\frac{ V^{(s)}_{pq}}{\lambda-iE_{pq}-i\omega s}.
\end{equation}
Notice, that evolving with the time independent Hamiltonian $H=H_0+W$, leads to the following perturbative expansion for the  effective $S$-matrix
\begin{equation}\label{flfl2}
[S_w^{(0)}]_{pq} = \frac{\delta_{pq}}{\lambda} ,\qquad [S_w^{(1)}]_{pq} = -\frac{i}{\lambda}\frac{W_{pq}}{\lambda-iE_{pq}}.
\end{equation}
The idea now, is that considering evolution only at the stroboscopic times we can transform \eqref{flfl} into \eqref{flfl2}
order by order, this way replacing periodic evolution by the quench evolution with a time independent potential.
We demonstrate how it is working in the first order, namely we present
\begin{equation}
S^{(1)}_{pq} = -i\sum_{s} V^{(s)}_{pq}\left(\frac{1}{\lambda-iE_{pq}-i\omega s}-\frac{1}{\lambda}\right) \frac{1}{iE_{pq}+i\omega s }.
\end{equation}
Then taking into account that inverse Laplace transform reads as $(\lambda+i\Omega)^{-1} \to e^{-i\Omega t}$ we see that for stroboscopic times we can ignore the denominators that are
proportional to $\omega$, this way
\begin{equation}
S^{(1)}_{pq} = -i\sum_{s}\frac{V^{(s)}_{pq}}{\lambda(\lambda-iE_{pq})} \frac{E_{pq}}{E_{pq}+\omega s } =-\frac{i}{\lambda}\frac{W_{pq}}{\lambda-iE_{pq}}  .
\end{equation}
From this we conclude that the effective potential in the first order reads as
\begin{equation}\label{W1 suppl}
W^{(1)}_{pq} = \sum_{s} V^{(s)}_{pq}\frac{E_{pq}}{E_{pq}+\omega s}.
\end{equation}
%The second order correction reads as
%\begin{equation}
%W^{(2)}_{pq} = \sum_{k,s_1,s_2} V^{(s_1)}_{pk}V^{(s_2)}_{kq}\left(
%\frac{E_{pk}}{E_{pk}+\omega s_1}-\frac{E_{pq}}{E_{pq}+\omega s_1+\omega s_2} \right) \frac{1}{E_{kq}+\omega s_2}.
%\end{equation}
We restrict our consideration here by this first order term. The full series for the effective potential can be also found, but we postpone its presentation for a separate publication.

\subsection{S5.2 Dynamics with the perturbative Floquet Hamiltonian}

Here we calculate perturbatively the averaged heating rate and current through the QPC in the time-periodic steady state established long after the onset of evolution. To this end we  calculate the increase of energy $\mathcal E$ and particle number in the right chain $N_R$ per period. Note that  $\mathcal E$ here is the energy of two chains without the QPC, i.e. the expectation value of the Hamiltonian $H_L+H_R$. The latter operator in general differs from $H_0$; however, the difference is local and bounded, therefore  the {\it  rate} of increase of expectation values of both operators in the steady state are identical.  It is essential for our calculations that $N_R$ commutes with  $H_L+H_R$. In general, our perturbative calculation is applicable to an arbitrary physical quantity $A$ that commutes with   $H_L+H_R$. We use the notation $A$ up to the end of calculations.

In this subsection we restrict ourselves to the harmonically driven $V_t=V e^{i\Omega t}+V^\dagger e^{-i\Omega t}+\overline V$, where the latter term is a time-independent contribution. Observe that $V$ need not be self-adjoint.  According to eq. \eqref{W1 suppl}, we obtain
\begin{equation}\label{W1 harmonic}
W^{(1)}_{pq}=V_{pq} \frac{E_{pq}}{E_{pq}+\omega }+ V_{qp}^* \frac{E_{pq}}{E_{pq}-\omega }+\overline V_{pq}.
\end{equation}

In leading order the evolution is governed by the Hamiltonian $H_{\rm F}^{(1)}=H_L+H_R+W^{(1)}$, and at large times it is sufficient to consider transition probabilities $P_{q \rightarrow k}(t)$ between the eigenstates $p$, $k$ of $H_L+H_R$ in the Fermi golden rule approximation,
\be
P_{q \rightarrow p}(t) \simeq t\, \left|W^{(1)}_{pq}\right|^2  \frac{\sin^2\left(t E_{pq}/2\right)}{t(E_{pq}/2)^2}.
\ee
Remember that we are interested exclusively in stroboscopic times $t=n\tau$, so that  $\sin^2\left(t E_{pq}/2\right)=\sin^2\left(t (E_{pq}\pm\omega)/2\right)$. Accounting for this leads to cancellation of the singularities of $W^{(1)}_{pq}$ at $E_{pq}=\pm \omega$ at stroboscopic times. Indeed, we substitute
\be
\frac{\sin^2\left(t E_{pq}/2\right)}{t(E_{pq}/2)^2(E_{pq}-\omega)^2(E_{pq}+\omega)^2} \rightarrow \frac{2\pi}{\omega^4}  \Big( \frac14\delta(E_{pq}+\omega)+\frac14\delta(E_{pq}-\omega)+\delta(E_{pq})\Big)
\ee
in the thermodynamic and large time limit. As a result, the increase of the observable $A$ is given by
\begin{multline}\label{Delta A(t)}
  \langle A \rangle_t-\langle A \rangle_0= \sum_{\eta, q} \rho_{\eta q} \sum_{\zeta,p} (A_{\zeta p}-A_{\eta q}) P_{\eta q \rightarrow \zeta p}(t) \\
  \rightarrow 8\pi t\,  \sum_{\zeta,\eta} \,\,\int\limits_{-1}^1 \frac{d\mathrm{E}_q}{2\pi\sqrt{1-\mathrm{E}_q^2}} \rho_\eta(\mathrm{E}_q) \int\limits_{-1}^1 \frac{dE_p}{2\pi\sqrt{1-E_p^2}}\,(A_{\zeta p}-A_{\eta q})
 \left|L \, \frac{E_{pq}^2-\omega^2}{\omega^2}\,W^{(1)}_{\zeta p; \eta q}\right|^2  \Big( \frac14\delta(E_{pq}+\omega)+\frac14\delta(E_{pq}-\omega)+\delta(E_{pq})\Big),
\end{multline}
where $A_{\zeta p}$ is a shorthand notation for the diagonal matrix element $A_{\zeta p;\zeta p}$.
Using \eqref{W1 harmonic}, we obtain
\begin{multline}\label{simpl}
 \left| \frac{E_{pq}^2-\omega^2}{\omega^2}\,W^{(1)}_{\zeta p; \eta q}\right|^2 \Big( \frac14\delta(E_{pq}+\omega)+\frac14\delta(E_{pq}-\omega)+\delta(E_{pq})\Big)= |V_{\zeta p; \eta q}|^2 \delta(E_{pq}+\omega)+|V_{\eta q;\zeta p}|^2 \delta(E_{pq}-\omega)+|\overline V_{\zeta p; \eta q}|^2 \delta(E_{pq}).
 \\
\end{multline}

To be even more specific, we choose
\begin{equation}\label{V via U and J}
J_t=(J e^{i\omega t}+J^* e^{-i\omega t})/2+\overline{J},\qquad \textrm{and}\qquad U^{L,R}_t=(U_{L,R}\, e^{i\omega t}+U_{L,R}^* \,e^{-i\omega t})/2+\overline{U}_{L,R}.
\end{equation}
Using eq. \eqref{eigenstates}, we get
\begin{equation}\label{Vpq explicit}
|L \, V_{\zeta p; \eta q}|= \sqrt{1-E_p^2} \, \sqrt{1-\mathrm{E}_q^2}\,\times
\left\{
\begin{array}{ll}
|U_L/2|, & \eta=\zeta=L,\\[0.5em]
|U_R/2|, & \eta=\zeta=R,\\[0.5em]
|J/2|, &\eta=L,\,\,\zeta=R \quad \textrm{or} \quad \eta=R,\,\,\zeta=L
%J/(2i), & \eta=R,\,\,\zeta=L,\\[0.5em]
%-J/(2i) , & \eta=L,\,\,\zeta=R
\end{array}
\right.
\end{equation}
and
\begin{equation}\label{overlineVpq explicit}
|L \, \overline V_{\zeta p; \eta q}|= \sqrt{1-E_p^2} \, \sqrt{1-\mathrm{E}_q^2}\,\times
\left\{
\begin{array}{ll}
| \overline U_L|, & \eta=\zeta=L,\\[0.5em]
| \overline U_R|, & \eta=\zeta=R,\\[0.5em]
| \overline J|, & \eta=L,\,\,\zeta=R \quad \textrm{or} \quad \eta=R,\,\,\zeta=L.
\end{array}
\right.
\end{equation}

At this point one could derive a general expression for $\partial_t\langle A\rangle_t$ from eqs. \eqref{Delta A(t)}, \eqref{simpl}, \eqref{Vpq explicit},\eqref{overlineVpq explicit}, which is, however, very bulky. Instead we turn specifically to the current and power.

We start from averaged current from left to right,
$
\overline{\mathcal{J}} = \lim\limits_{n\rightarrow\infty}(\langle N_R\rangle_{t_n}-\langle N_R\rangle_{t_{n-1}})/\tau.
$
Substituting $A=N_R$, we obtain
 \begin{equation}
A_{\zeta p}-A_{\eta q}=\left\{
\begin{array}{ll}
1, & \eta=L,\,\,\zeta=R, \\[0.5em]
-1, & \eta=R,\,\,\zeta=L, \\[0.5em]
0, & \eta=\zeta,
\end{array}
\right.
 \end{equation}
which entails
\begin{equation}\label{current}
\overline{\mathcal{J}}^{(1)} = \int\limits_{-1}^1 \frac{dE}{2\pi}\sqrt{1-E^2}\, \left(\rho_L-\rho_R\right) \left(|J|^2\, {\rm Re}\left[\sqrt{1-(E+\omega)^2}+\sqrt{1-(E-\omega)^2}\right] + 4 \,|\overline J|^2 \, \sqrt{1-E^2} \right),
\end{equation}
where we drop the argument in $\rho_{L,R}(E)$ for brevity. This leads to eqs. (15),(17) of the main text.

Now we turn to calculating the power $\overline{\mathscr{W}} = \lim\limits_{n\rightarrow\infty}(\langle H_L+H_R\rangle_{t_n}-\langle H_L+H_R\rangle_{t_{n-1}})/\tau$.
In this case  $A=H_L+H_R$ and $A_{\zeta p}-A_{\eta q}=E_{pq}$. This leads to
\begin{align}\label{power}
\overline{\mathscr{W}}^{(1)} & = \omega \int\limits_{-1}^1 \frac{d\mathrm{E}_q}{2\pi}\sqrt{1-E^2}
 \left(\rho_L(|J|^2 +|U_L|^2) +\rho_R(|J|^2 +|U_R|^2)\right) {\rm Re}\left[\sqrt{1-(E+\omega)^2}-\sqrt{1-(E-\omega)^2}\right].
\end{align}
This way we get eqs. (14),(16) of the main text.

\section{S6. Numerical analysis}

We numerically simulate the dynamics of finite systems with $L=50$ sites in each chain. The maximal group velocity in the tight-binding chain with the Hamiltonian (2) is equal to $1$, therefore the local state in the vicinity of the QPC starts to be affected by finite size effects (namely, by the reflection of excitations from the boundaries of the chains) at $t=2L$. To estimate the heating rate and the current in the large time limit, we consider the last driving cycle before these finite-size distortions come into play.

\begin{figure}[t] %  figure placement: here, top, bottom, or page
		\centering
		\includegraphics[width=0.9 \linewidth]{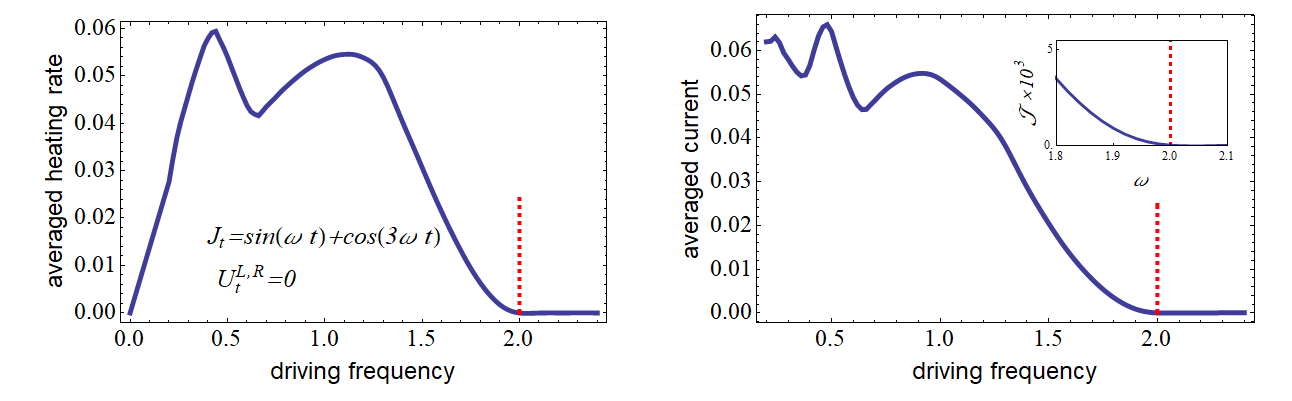}\\
        \includegraphics[width=0.9 \linewidth]{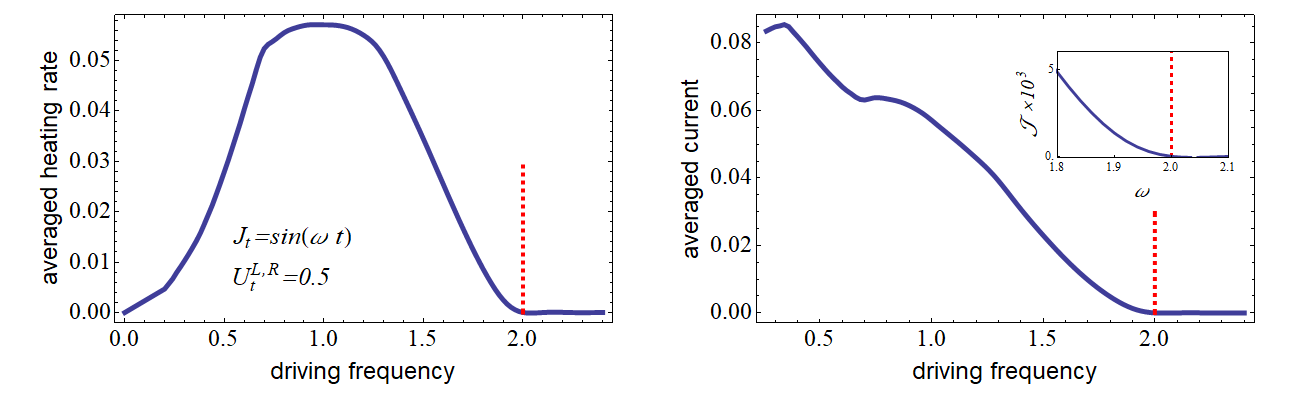}\\
		\includegraphics[width=0.9\linewidth]{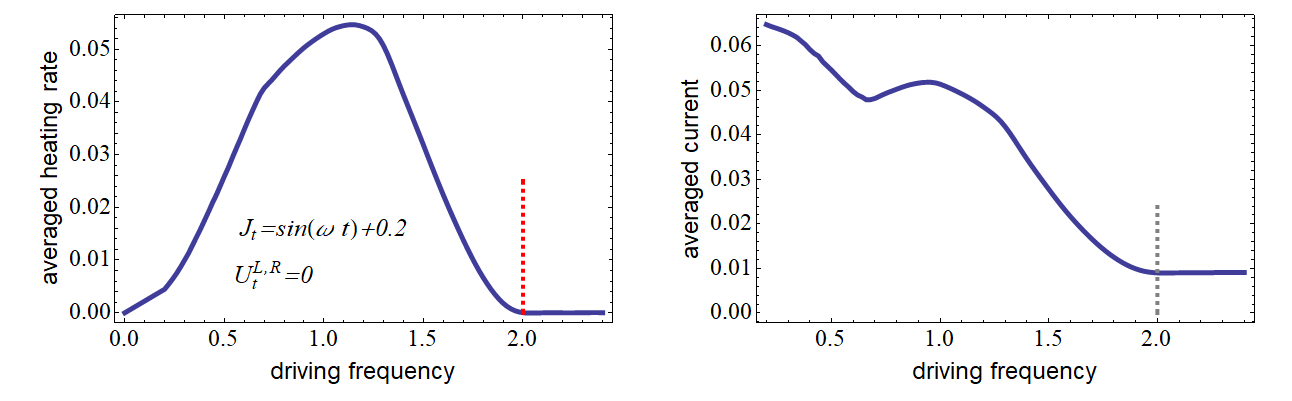}\\
        \caption{Average heating rate  $\overline{\mathscr{W}}$  and average current through the QPC $\overline{\mathcal{J}}$ for various QPCs with constant on-site potentials. One can see that in this case the current vanishes above $\omega=2$ whenever the average tunneling amplitude is zero. The vicinity of $\omega=2$ is zoomed where necessary. }
		\label{Fig:numerical 1}
\end{figure}

\begin{figure}[t] %  figure placement: here, top, bottom, or page
		\centering
		\includegraphics[width=0.9\linewidth]{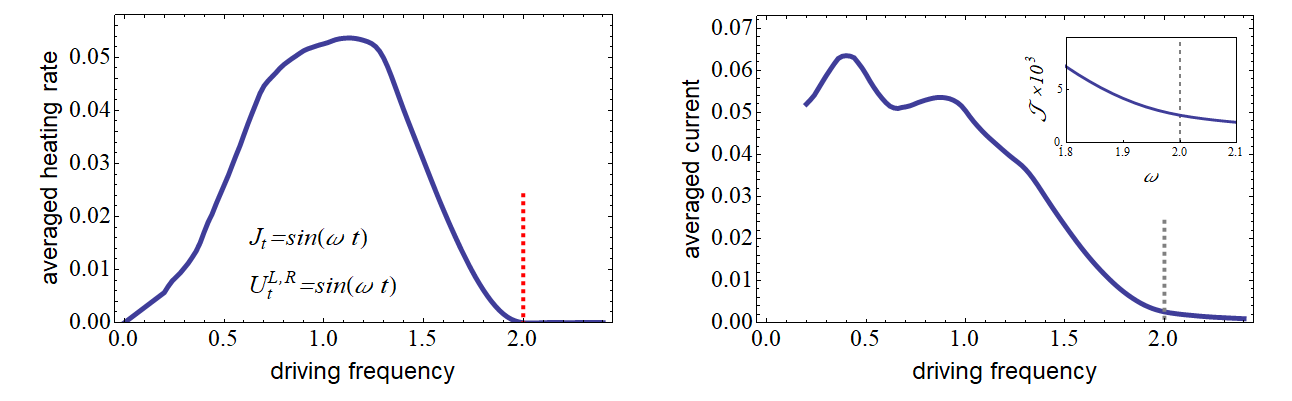}\\
        \includegraphics[width=0.9\linewidth]{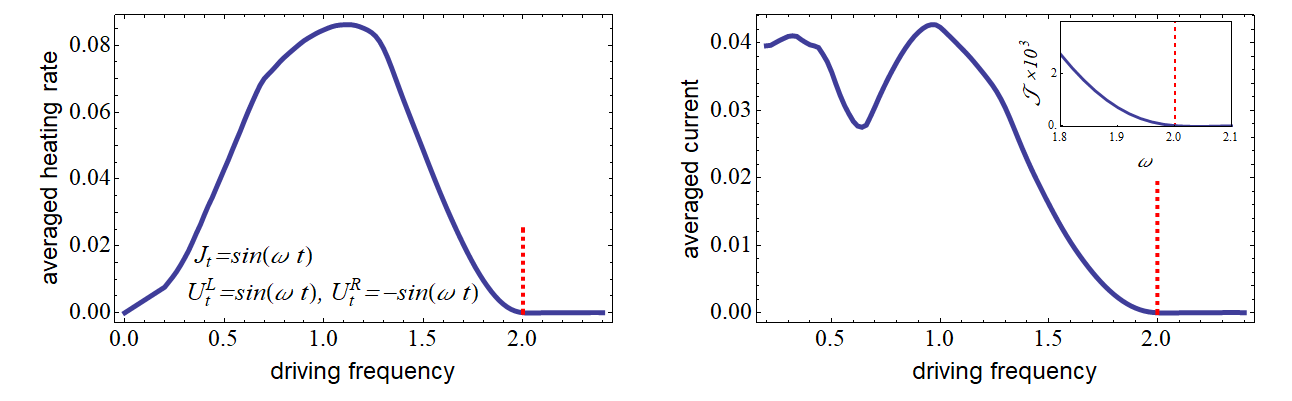}\\
          \caption{Average heating rate  $\overline{\mathscr{W}}$  and average current through the QPC $\overline{\mathcal{J}}$ for two QPCs with time-dependent on-site potentials.   The vicinity of $\omega=2$ is zoomed in the left panels. }
		\label{Fig:numerical 2}
\end{figure}

\begin{figure}[t] %  figure placement: here, top, bottom, or page
		\centering
        \includegraphics[width=0.9\linewidth]{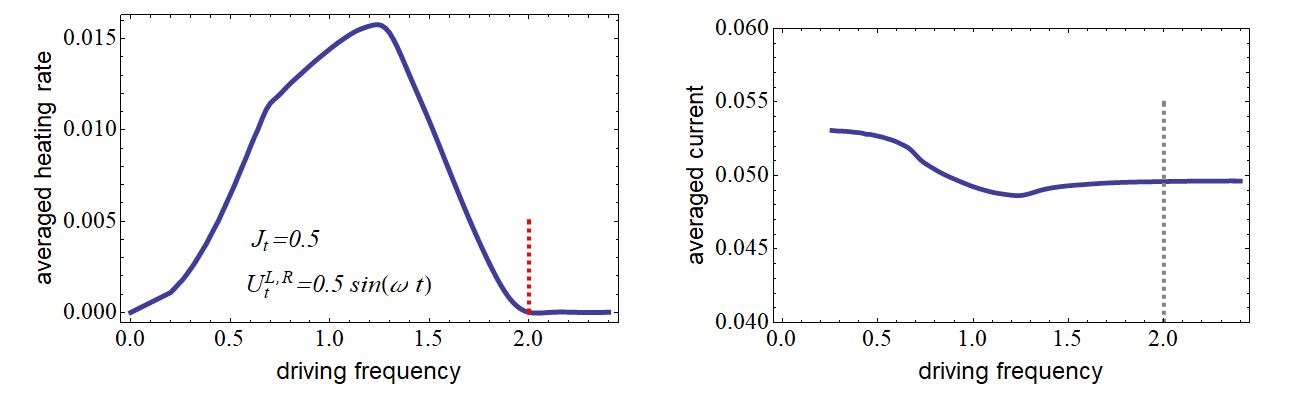}\\
        \caption{Average heating rate  $\overline{\mathscr{W}}$  and average current through the QPC $\overline{\mathcal{J}}$ for a QPC with a constant tunneling amplitude.  }
		\label{Fig:numerical 3}
\end{figure}

We present average heating rate  $\overline{\mathscr{W}}$  and average current through the QPC $\overline{\mathcal{J}}$ for various QPCs in Figs. \ref{Fig:numerical 1}-\ref{Fig:numerical 3}. The type of QPC is indicated in each row of these figures by specifying the functions $J_t$, $U_t^{L,R}$.  For all these plots we consider systems with $N=20$ fermions initially residing in the left chain at zero temperature, the right chain being empty.

One can see that $\overline{\mathscr{W}}$ vanishes for $\omega\geq 2$ for all QPCs considered. As for the current, the situation is more complicated.

Fig. \ref{Fig:numerical 1} illustrates that if the on-site potentials are constant, the current experiences phase transition at $\omega=2$ and admits a value independent on $\omega $ for $\omega>2$. This value is zero if the averaged tunneling amplitude is zero.

In Fig. \ref{Fig:numerical 2} we present two very similar QPCs with time-dependent on-site potentials. One of them features the vanishing of the current while another does not.

In Fig. \ref{Fig:numerical 3} we present a case of constant tunneling amplitude. There the current does not vanish. This highlights the importance of driving  the tunneling amplitude to halt the current.

Note that the driving need not be harmonic for our conclusions to hold. This is illustrated by the upper row of Fig.~\ref{Fig:numerical 1}.

Finally, we present an animation of the real time dynamics of the particle density profile in the supplementary file \texttt{cartoonL100N60.gif}. There $\omega=2.2$, $L=100$ and initially the left chain contains $N=60$ particles in the ground state while the right chain is empty. Two QPCs are considered - the conformal QPC (eq. (4) of the main text) and the tunneling QPC (eq. (11) of the main text). One can see that in the former case a steady state current is established, while in the latter case the current is halted after an initial transient.

\section{S7. Merits of the Floquet-Magnus expansion}

In this section we explore the merits of the Floquet-Magnus  expansion (FME) applied to the problem at hand. The FM expansion is a formal expansion of the Floquet Hamiltonian in powers of $1/\omega$~\cite{kuwahara2016floquet-magnus},
\be
H_{\rm F}=\sum_{n=0}^\infty \omega^{-n} M^{(n)}.
\ee
The first three terms of the  FME  read
\begin{align}\label{FM}
M^{(0)}= &\tau^{-1} \int_0^\tau H_{t_1} dt_1,\\
M^{(1)}= &\frac{i\omega}{2} \int_0^\tau dt_1 \int_0^{t_1} dt_2 \, [H_{t_1},H_{t_2}],\\
M^{(2)}= &\frac{\omega^2}{6} \int_0^\tau dt_1 \int_0^{t_1} dt_2  \int_0^{t_2}dt_3 \, \Big([H_{t_1},[H_{t_2},H_{t_3}]+ [H_{t_3},[H_{t_2},H_{t_1}], \Big), \label{FM2}
\end{align}
see e.g. \cite{kuwahara2016floquet-magnus}.
The higher terms have analogous structure of nested commutators.

Convergence of the FME can be proven for frequencies greater than $c\, \tau^{-1} \int_0^{\tau} dt ||H_t||$, where $||H_t||$ is the operator norm and $c$ is some numerical constant \cite{blanes2009magnus,kuwahara2016floquet-magnus}. The norm of a many-body Hamiltonian diverges in the thermodynamic limit, therefore it is widely believed that the FME has, in general, a zero convergence radius for many-body systems (however, the truncated FME can still be used to describe a transient dynamics at finite times \cite{Bukov_2015,kuwahara2016floquet-magnus}).

The case of noniteracting fermions is somewhat subtle. One could argue that the above-mentioned convergence result could be applied to a one-body problem, where the corresponding operator norm is finite \cite{Bukov_2015}. While this is indeed the case, this does not guarantee the uniform convergence of FME for a corresponding many-body noninteracting Hamiltonian in the thermodynamic limit. This is because subleading terms in a one-body Hamiltonian, that vanish in the thermodynamic limit, can add up to a finite (and even divergent) term in the corresponding many-body Hamiltonian. In particular, in the case of a local driving, the heating rate can vanish in the thermodynamic limit as $O(1/L)$ for a one-body problem, but maintain a finite value for a many-body problem.

Thanks to the exact solution for the conformal QPC, we are able to explicate the inapplicability of the FME in this case. Substituting the Hamiltonian (2)-(4) to eqs. \eqref{FM}-\eqref{FM2}, we obtain
\begin{align}\label{M0}
 M^{(0)}=&H_L+H_R,\\
  M^{(1)}= & \frac{i}4 \left(c^\dagger_{L-1} \, c^\dagger_L \, c_{L+1}^\dagger \, c_{L+2}^\dagger \right)
  \left(
\begin{array}{cccc}
0 & 0 &-1 & 0\\
0 & 0 & 1& 1\\
1& -1 & 0 & 0 \\
0& -1& 0 & 0 \\
\end{array}\right)\left(
\begin{array}{c}
c_{L-1}\\c_L \\ c_{L+1}\\ c_{L+2}
\end{array}
\right), \\
  M^{(2)}= & \frac{1}8 \left(c^\dagger_{L-2} \,c^\dagger_{L-1} \, c^\dagger_L \, c_{L+1}^\dagger \, c_{L+2}^\dagger\, c_{L+3}^\dagger \right)
  \left(
\begin{array}{cccccc}
0 & 0 &1 & 0 & 0 & 0\\
0 & -2 & 1 & 0 & 0 & 0\\
1 & 1 & 0 & 0 & 0 & 0\\
0 & 0 & 0& 0 & 1 & -1 \\
0 & 0 & 0& 1& 2 & 0 \\
0 & 0 & 0 & -1 & 0 & 0
\end{array}\right)\left(
\begin{array}{c}
c_{L-2}\\c_{L-1}\\c_L \\ c_{L+1}\\ c_{L+2} \\ c_{L+3}\\
\end{array}
\right),\\
.~.~.& \nonumber
\end{align}

One can see that the FME truncated at the order $n$ contains hopping terms over at most $n$ sites. At the same time,  the exact Floquet Hamiltonian
%\eqref{HFl}
(7)  contains hoppings over the entire system, i.e. over $L$ sites.

As a result of the local nature of hoppings in the truncated FME for any finite  $n$, the heating rate, calculated with the truncated FME instead of the true Floquet Hamiltonian, vanishes irrespective of the value of $\omega$. Thus it is impossible to derive the phase transition between the heating and the no-heating regime by means of the FME truncated at a finite order.

\end{document}